\documentclass{article}

\usepackage{arxiv}

\usepackage[utf8]{inputenc} 
\usepackage[T1]{fontenc}    
\usepackage{hyperref}       
\usepackage{url}            
\usepackage{booktabs,longtable}       
\usepackage{siunitx}
\usepackage{amsfonts}       
\usepackage{nicefrac}       
\usepackage{microtype}      
\usepackage{lipsum}
\usepackage{xcolor}
\usepackage{float}

\usepackage{mathtools}
\usepackage{float}
\usepackage[style=nature,backend=biber,doi=false,url=false]{biblatex}
\title{Homogeneous Nucleation of Sheared Liquids: Advances and Insights from Simulations and Theory}

\author{
	\href{https://orcid.org/0000-0001-8706-2383}{\includegraphics[scale=0.06]{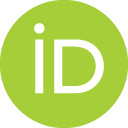}\hspace{1mm}Amrita Goswami}\\
  Department of Chemical Engineering\\
  Indian Institute of Technology Kanpur\\
  \texttt{amritag@iitk.ac.in} \\
	\And
	\href{https://orcid.org/0000-0001-8056-2115}{\includegraphics[scale=0.06]{orcid.png}\hspace{1mm}Jayant K. Singh}\footnotemark[1] \\
	Department of Chemical Engineering\\
	Indian Institute of Technology Kanpur\\
	\texttt{jayantks@iitk.ac.in} \\
}

\begin{document}
\maketitle

\begin{abstract}
	One of the most ubiquitous and technologically important phenomena in
  nature is the nucleation of homogeneous flowing systems. The
  microscopic effects of shear on a nucleating system are still
  imperfectly understood, although in recent years a consistent picture
  has emerged. The opposing effects of shear can be split into two major
  contributions for simple liquids: increase of the energetic cost of
  nucleation, and enhancement of the kinetics. In this review, we
  describe the latest computational and theoretical techniques which
  have been developed over the past two decades. We collate and unify
  the overarching influences of shear, temperature, and supersaturation
  on the process of homogeneous nucleation. Experimental techniques and
  capabilities are discussed, against the backdrop of results from
  simulations and theory. Although we primarily focus on simple liquids,
  we also touch upon the sheared nucleation of more complex systems,
  including glasses and polymer melts. We speculate on the promising
  directions and possible advances that could come to fruition in the
  future.
\end{abstract}


\newcommand{\Tau}{\mathrm{T}}


\section{\label{introduction}Introduction}

Understanding quiescent homogeneous nucleation has been a longstanding
goal of the academic community, which is justifiable considering its
immense importance in various scientific and industrial applications
\cite{Chayen2006, KeltonGreer2010, Lee2011, BartelsRausch2013, Murray2010a, Zhang2018, Murray2012}.
Quiescent homogeneous nucleation is opaque to interpretations across the
spectrum of modern techniques, from both theoretical and experimental
viewpoints. The process of quiescent homogeneous nucleation is notoriously
challenging to study using computational methods, since it is a
stochastic rare event in terms of the time and length scales. On the
other side of the spectrum, nucleation experiments are plagued by rapid
crystallization or vitrification \cite{Tanaka2019} which can be
difficult to capture microscopically \cite{Sosso2016a}. A phenomenon of
perhaps even greater theoretical and technological relevance
\cite{Penkova2006, Woodhouse2013, Berland1992, Baird2011, Koh2011} is
the nucleation of flowing liquids---in reality, liquids are rarely
static.

The existing literature is somewhat divided on the topic of the effect
of shear and its influence on nucleation. Certain results indicate that
shear impedes and inhibits nucleation \cite{Blaak2004, Blaak2004a},
while others suggest that flow induces or enhances nucleation
\cite{Mokshin2009, Graham2009, Radu2014, Forsyth2014, Sharma2014, Shao2015, Stroobants2020}.
These seemingly conflicting results can be reconciled within a broader
context, bolstered by experimental \cite{Holmqvist2005, Liu2013} and
computational
\cite{Allen2008, Allen2008a, Cerda2008, Lander2013, Mura2016, Mokshin2010, Mokshin2013, Peng2017, Luo2020, Goswami2020a, Goswami2021a}
studies which imply that the nucleation rate is non-monotonic with
shear. A linear flow field tends to deform incipient clusters. This
shear deformation affects the energetics of the nucleation process. At
the same time, an imposed shear rate encourages the diffusive motion of
the particles in a simple liquid, tending to enhance the kinetics of
nucleation. The ordering of polymer strands can also be facilitated by
shear. A detailed and unified evaluation of these underlying mechanisms forms the
kernel of this review.

The study of sheared homogeneous nucleation, particularly of simple
liquids, glassy systems and colloids, via theory and simulations is a
burgeoning field. Theoretical and simulation studies of sheared
nucleation have experienced something of renaissance in the past decade,
partly due to the surge in computational power \cite{Shalf2020}. We elucidate recent
theoretical developments and advances in the field, focusing mostly on
simple liquids. This paper is organized as follows: 
in the remaining
introductory sections, we provide a brief overview of Classical
Nucleation Theory (Section \ref{overview-of-conventional-cnt}) and the
seeding method (Section \ref{cnt-and-the-seeding-method}) for quiescent
homogeneous nucleation. Section \ref{computational-strategies-for-sheared-nucleation} is dedicated to
computational strategies which can be used for sheared homogeneous
nucleation, describing brute-force approaches (Section
\ref{brute-force-approaches}), forward-flux sampling (Section
\ref{forward-flux-sampling}), and recent CNT-based approaches (Section
\ref{classical-nucleation-theory-approaches}). In Section
\ref{results-from-simulations-and-theory} we expound the various effects
of shear on homogeneous nucleation behaviour obtained from simulations
and theory, and explore microscopic mechanisms. Section
\ref{experiments} covers experiments, emphasizing current capabilities
of typical experimental techniques. Finally, in Section
\ref{future-scope-and-challenges}, we summarize the major conclusions,
highlight exciting possible extensions of the field, and identify future
challenges.

\subsection{\label{overview-of-conventional-cnt}Overview of Conventional
CNT}

\begin{figure}[H]
 \centering
 \includegraphics[width=0.7\textwidth]{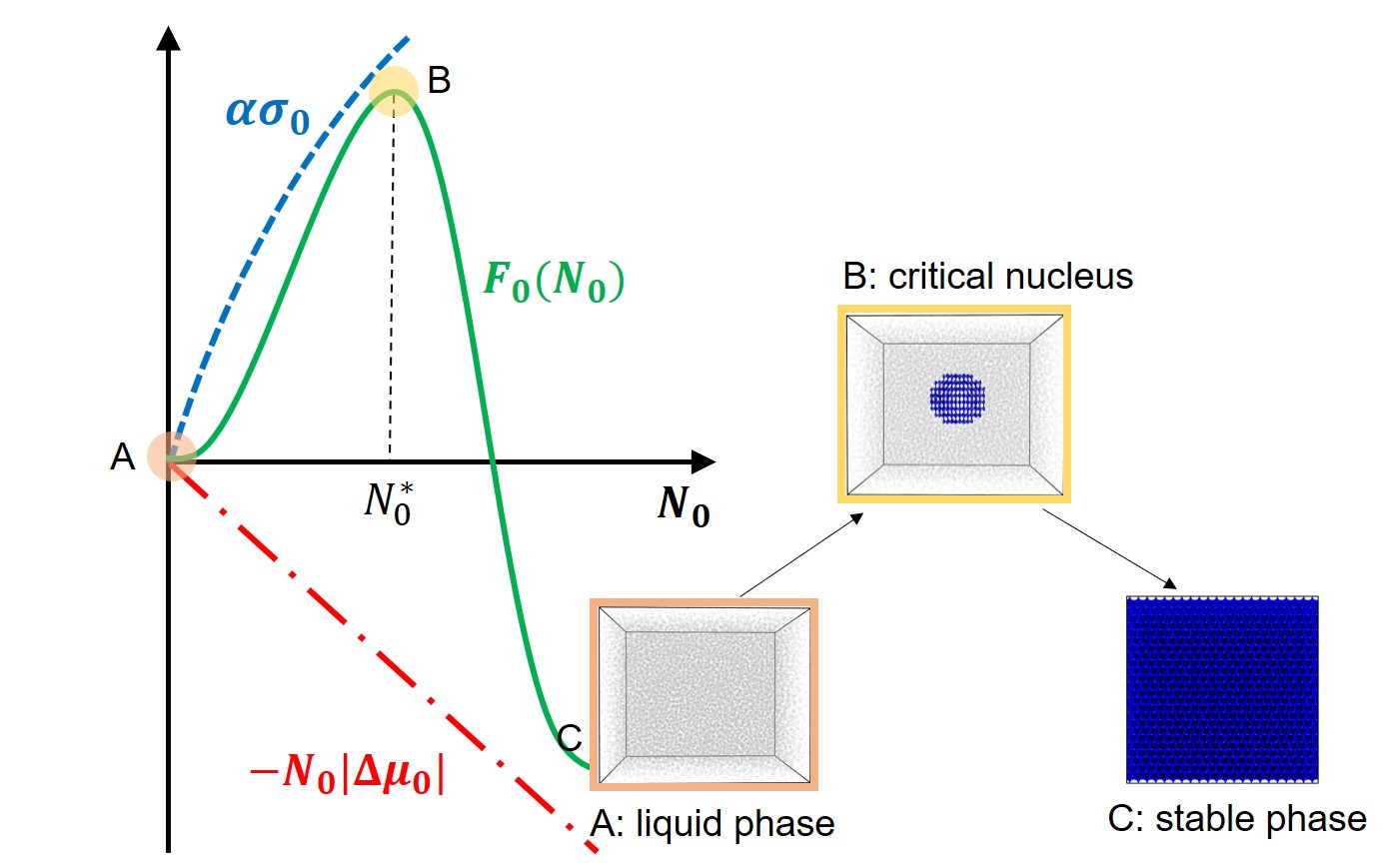}
 \caption{Schematic of the free energy curve for homogeneous nucleation, $F_0(N_0)$. The favourable volume contribution to the free energy, $\alpha \sigma_0$, and unfavourable interfacial free energy cost, $-N_0 |\Delta \mu_0|$, are indicated with blue and red dashed lines, respectively. Insets depict various theoretical stages of the nucleation process, from the initial state (A) to the final stable phase (C).}
 \label{fig:CNTfreeEnergN-noShear}
\end{figure}

Classical Nucleation Theory (CNT) is an approximate phenomenological
theory, whose ubiquity cannot be overstated even though about \(100\)
years have passed since its conception. Aspects of CNT are ingrained some
form in virtually every computer simulation of homogeneous nucleation of
simple liquids. Exhaustive treatments and reviews of CNT are available elsewhere in the literature \cite{Kelton1991, Oxtoby1992, KeltonGreer2010, Ickes2015}.
Here, we summarize details of relevance for further extensions to sheared homogeneous nucleation.  

Originally formulated for describing the condensation of a vapour into a
liquid, CNT has since been found to be applicable to other phase
transitions. In particular, extensions to CNT have dominated
descriptions of crystal nucleation of supercooled and supersaturated
solutions. The cornerstone of CNT is the assumption that microscopic
crystalline nuclei can be effectively treated as scaled-down replicas of
the macroscopic crystalline phase. These crystalline clusters are
considered to be separated from the metastable phase by a "dividing
surface" of negligible thickness. This negligibly thin and sharp
"dividing surface" is a convenient construct and not a statement about
the molecules and their physical reality. Furthermore, CNT assumes that
crystal nucleation is a single-step process, wherein the only
significant barrier is the free energy barrier arising from the
interplay of the solid-liquid interfacial energy, \(\sigma_0\), and the
chemical potential difference, $\Delta \mu_0$. All other
contributions to the free energy are neglected.

With these assumptions, in the archetypal CNT formulation, the free
energy of formation \(F_0\) of a spherical crystallite nucleus,
containing \(N_0\) particles, is expressed as the sum of a favourable
volume term and an unfavourable surface term
\cite{Volmer1926, Becker1935, Kelton1991, Sunyaev1992, Sosso2016a}:

\begin{equation}\label{eqn:Gibbscnt1}
  F_0(N_0) = \underbrace{-N_0 |\Delta \mu_0|}_{\textrm{volume term}} + \underbrace{\alpha \sigma_0}_{\textrm{surface term}},
\end{equation}

where $\Delta \mu_0$ is the chemical potential difference between
the metastable liquid phase and the crystal, \(\sigma_0\) is the
crystal-fluid interfacial free energy,
\(\alpha = (36 \pi N_0^2 v'^2)^{\frac{1}{3}}\) is the geometric shape
factor of a spherical nucleus, and \(v'\) is the volume of one unit
particle or molecule in the crystal. In order to emphasize the fact that
quintessential CNT is used for quiescent systems not under the influence
of shear, the subscript $0$ has been used for the concerned variables. For example, the system is assumed to be quiescent in the calculation of the quantities $F_0$, $N_0$, $\Delta \mu_0$ and $\sigma_0$ in Eq. \eqref{eqn:Gibbscnt1}. This convention will
be adopted throughout this review to refer to parameters that are estimated in the absence of flow.

In Eq. \eqref{eqn:Gibbscnt1}, \(N_0 |\Delta \mu_0|\) is the driving
force for the nucleation prcoess, arising from the greater thermodynamic
stability of the solid phase, compared to the metastable liquid phase
\cite{Kelton1991, Auer2004a}. The free energy cost of forming the
hypothetical interface is given by \(\alpha \sigma_0\). The competition
between the favourable and unfavourable terms yields a maximum in
\(F_0(N_0)\), indicative of a kinetic barrier to the phase
transformation, depicted in Figure \ref{fig:CNTfreeEnergN-noShear}.


Crystalline clusters form spontaneously via stochastic, localized
fluctuations. A physical interpretation of Eq. \eqref{eqn:Gibbscnt1} is
that the unfavourable surface term dominates for small nuclei, which
consequently tend to dissipate into the liquid phase unless they are
aided by fortuitous fluctuations. A cluster grows spontaneously,
overcoming the free energy barrier, only if it exceeds a critical size
\cite{Auer2005}. The CNT expression for this free energy barrier for
nucleation is given by:

\begin{equation}\label{eqn:chap3cntFreeEnerg}
  F_0(N_0^*) = \frac{N_0^* |\Delta \mu_0 |}{2} = \frac{16 \pi v'^2 \sigma_0^3}{3 |\Delta \mu_0 |^2},
\end{equation}

where \(F_0(N_0^*)\) is the height of the free energy barrier, and
\(N_0^*\) is the number of particles or molecules in the critical
cluster. Thus, the stochastic regime, preceding the formation of the
critical nucleus, gives way to crystal growth, which is a deterministic
thermodynamically driven regime.

Assuming that crystalline nuclei grow or shrink due to the attachment or
dissolution of atoms (or `monomers') one-by-one the steady-state rate of
nucleation can be estimated from an Arrhenius reaction rate type
equation: \cite{Kelton1983, Kelton1991, Auer2001, Auer2004}

\begin{equation}\label{eqn:chap3rateCNT}
  J_0 = \rho_l Z_0 f_0^+ e^{-\frac{F_0(N_0^*)}{k_B T}},
\end{equation}

where \(\rho_l\) is the number density of the supercooled liquid,
\(Z_0\) is the Zeldovich factor \cite{Kelton1991}, and \(f_0^+\) is the
rate of attachment of molecules to the cluster in units of inverse time.
\(k_B\) is the Boltzmann constant, and \(T\) is the temperature. The
Zeldovich factor, \(Z_0\), captures the multiple possible re-crossings
of the free energy barrier \cite{Pan2004} and is related to the
curvature of the free energy curve at the critical cluster size
\(N_0^*\).

The nucleation rate is thus proportional to the thermodynamic
probability of a fluctuation yielding the formation of a critical
cluster, and a dynamical factor \((\rho_lZ_0 f_0^+)\), called the
kinetic prefactor. Since \(f_0^+\) is related to the time required for a
single particle or molecule to attach itself to the solid cluster, it
can expressed as \cite{Kelton1983, Kelton1991}:

\begin{equation}\label{eqn:chap3diffusion2}
  f_0^+ = \frac{24 D_0 (N_0^*)^{\frac{2}{3}}}{\lambda^2},
\end{equation}

where \(D_0\) is the bulk self-diffusion coefficient of the supercooled
liquid phase at a particular temperature \(T\), \(\lambda\) is the
atomic `jump length', estimated to be about one molecule diameter.

Therefore, CNT provides a convenient theoretical framework for
calculating the free energies (Eq. \ref{eqn:chap3cntFreeEnerg}) and
nucleation rates (Eq. \ref{eqn:chap3rateCNT}). However, CNT involves
several simplifying assumptions, which we enumerate and summarize
briefly, revealing the shortcomings of the theory.

\begin{enumerate}
\def\labelenumi{\arabic{enumi}.}
\item
  Conventional CNT assumes that nucleation proceeds according to a
  single-step mechanism, via which clusters of the ordered phase form
  directly from the metastable parent phase. However, there is
  considerable theoretical and experimental evidence for a two-step
  model of nucleation for proteins \cite{Nicolis2003, Pan2005, Talanquer1998, Sauter2015, Vorontsova2015, Russo2016, Toernquist2018, Kellermeier2012} and supersaturated solutions \cite{Shore2000, Pontoni2004, Gavezzotti1999, Salvalaglio2015, Erdemir2007, Vehkamaeki2006, Parveen2005},
  which asserts that a cluster of monomers forms first, followed by
  reorganization into the ordered phase, i.e., the cluster size and
  order parameter are decoupled. Meticulous reviews in the literature
  have covered two-step and non-classical nucleation in detail
  \cite{Jin2020, Sleutel2018, Erdemir2009, Vekilov2010, Sear2012, Yoreo2013, Gebauer2014, Zahn2015, Zaccaro2001, Karthika2016, Zhang2014}.
  In this review, we focus on single-component systems which can be
  reasonably presumed to follow a single-step mechanism.
\item
  The slow kinetics originate solely from the free energy barrier, and
  microscopic motions are fast compared to the time-scales of
  nucleation. In other words, only the
  slower degrees of freedom contribute to the free energy barrier. The
  size of the crystal embryo changes via the attachment or detachment of
  monomers one at a time \cite{Sear2012}. Where the microscopic motions
  show a tendency towards colloidal aggregation or secondary structure,
  the CNT model fails. In particular, conventional CNT typically cannot
  be applied without further extensions to glassy liquids and polymer
  systems, which are prone to ageing and `slow' rearrangement of
  molecules or chains \cite{Xu2021, Abyzov2020}.
\item
  In quintessential CNT, the nucleus is assumed to be perfectly
  spherical. This has been shown to be a reasonable approximation for
  water \cite{Espinosa2014, Espinosa2016}. CNT has also been extended
  for ellipsoidal clusters \cite{Lutsko2018}.
\item
  The embryos are assumed to be perfectly identical to the bulk ice
  phase, and are thus conceptualized as fragments of the bulk phase.
  Finite size effects on the physical and structural properties of the
  nuclei are wholly neglected.
\item
  The solid-liquid interface is considered to be a sharp dividing
  surface. However, in reality, the interfacial region is finite, and
  tends to be a few molecules thick. This approximation is more
  glaringly inaccurate for smaller nuclei wherein the interfacial
  molecules form a significant proportion.
\item
  The microscopic description of the interfacial energy \(\sigma_0\), is
  only theoretically defined for a hypothetical isotropic nucleus (an
  assumption which often falls under the purview of the ideal
  `capillary approximation'), and can be ambiguous for real systems
  \cite{Auer2004a, Ickes2015}. In early CNT formulations, the solid-liquid
  interfacial energy, \(\sigma_0\), at any temperature was assumed to be
  constant and equal to the macroscopic interfacial energy at the
  equilibrium freezing temperature \cite{Oxtoby1992}. This approximation
  presumes that neither the curvature of nucleus nor the temperature
  has an effect on \(\sigma_0\). However, subsequent extensions to CNT account
  for the effect of the nucleus size, curvature and supercooling in the
  estimation of \(\sigma_0\). These modified CNT-based techniques include the so-called `seeding
  method' \cite{Bai2005, Knott2012, Sanz2013} and a formulation invoking
  additional corrections for fuzzy solid-liquid interfaces and
  non-spherical cluster shapes \cite{Prestipino2012}. We elaborate the
  seeding class of techniques in this review, and demonstrate how they
  can be employed for sheared liquids.
\end{enumerate}

Despite these inherent caveats, CNT has been shown to be surprisingly
accurate \cite{Bai2005}, and is often the pragmatic technique of choice
due to its simplicity and wide applicability. CNT results are also used
as a benchmark for comparing and discussing qualitative trends with more
sophisticated approaches \cite{HajiAkbari2015, Jiang2018}. In the
remaining sections, we show that CNT provides a convenient basis for
describing the nucleation of sheared supercooled liquids. Nevertheless,
the gross simplifying assumptions make CNT far from a universal panacea.

\subsection{\label{cnt-and-the-seeding-method} CNT and the Seeding
Method}

\begin{figure}[H]
\centering
\includegraphics[width=\textwidth]{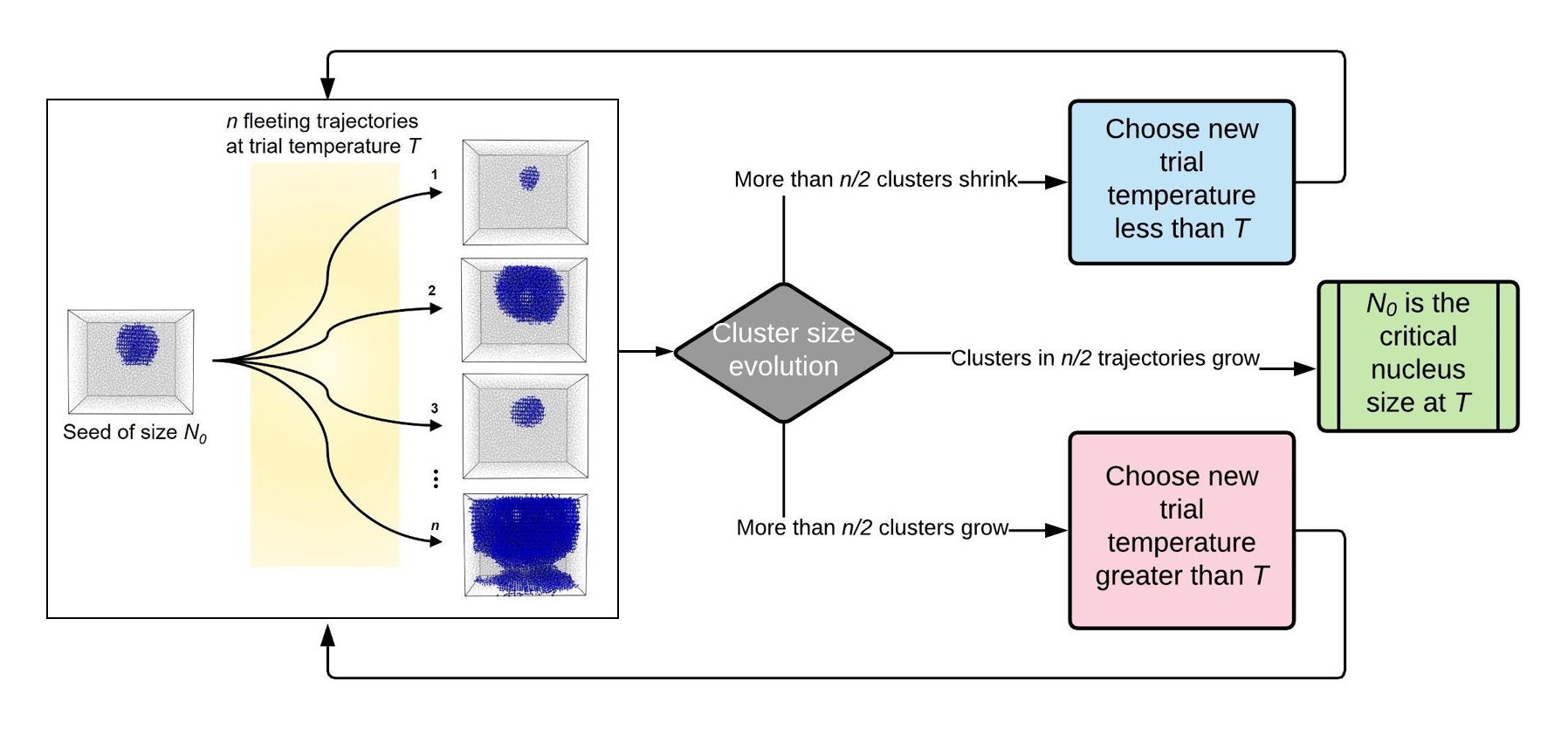}
\caption[Flow chart of a typical implementation of the seeding technique.]{Flow chart of a typical implementation of the seeding technique. The inserted seed size, $N_0$, is known. Collections of $n$ trajectories are run at trial temperatures iteratively, until a temperature is found at which clusters in about $n/2$ trajectories enlarge, while those in the remaining $n/2$ trajectories decrease in size. Alternatively, if the critical nucleus size at a specific value of $T$ is desired, the seed size, $N_0$, is iteratively modified until the stopping criterion is satisfied.}
\label{fig:seedingFlowChart}
\end{figure}


Perhaps the most egregious of the CNT assumptions is the capillary
approximation. Elements of this approximation can be dismantled by
estimating the interfacial energy, \(\sigma_0\), as a function of the
critical cluster curvature and supercooling. A popular technique for
such calculations of \(\sigma_0\) is the seeding method, which is an
approximate simulation approach used within the framework of CNT
\cite{Bai2005, Knott2012, Sanz2013}.

\begin{figure}[H]
\centering
\includegraphics[width=0.7\textwidth]{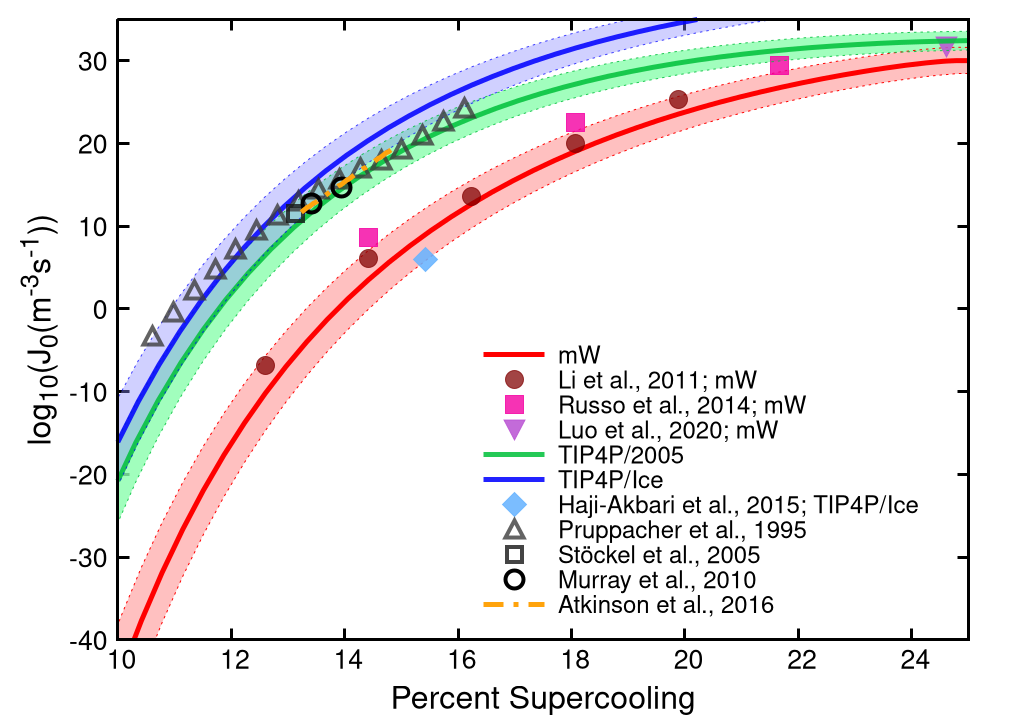}
\caption{Comparison of the homogeneous nucleation rates, $J_0$, as a function of the percent supercooling, estimated using the seeding method (solid lines) for the mW, TIP4P/2005 and TIP4P/Ice water models, with results obtained from experiments (open symbols and dashed line) and other simulation strategies (filled symbols). The shaded regions flanking the solid lines denote the errors associated with seeding calculations \cite{Espinosa2014}. $J_0$ was calculated via seeding, using data from Espinosa et al. \cite{Espinosa2014} and power law fits to the diffusion coefficients. Results for the mW model are denoted by filled maroon circles, magenta squares, and an inverted mauve triangle, obtained via FFS calculations by Li et al. \cite{Li2011}, a modified umbrella sampling scheme by Russo et al. \cite{Russo2014}, and MFPT calculations by Luo et al. \cite{Luo2020}, respectively. The filled blue diamond represents the rate for the TIP4P/Ice model estimated via FFS, at $230 \ K$ by Haji-Akbari et al \cite{HajiAkbari2015}. Experimental rate estimates were reported by Pruppacher et al \cite{Pruppacher1995}. (open grey triangles), Stöckel et al \cite{Stoeckel2005}. (open grey squares), Murray et al \cite{Murray2010}. (open black cicles), and Atkinson et al \cite{Atkinson2016} (mustard dashed line).}
\label{fig:seedingWatNoShear}
\end{figure}

The seeding method, which was first proposed by Bai et al
\cite{Bai2005, Bai2006a}, can reliably determine details of the critical
nucleus, at a particular condition of metastability. When used in tandem
with CNT, the seeding technique is capable of estimating interfacial
free energies from microscopic data, along with the free energies and
rates for conditions of shallow metastability, which are generally
inaccessible to brute-force approaches
\cite{Pereyra2011, Knott2012, Sanz2013, Espinosa2014, Espinosa2015, Zimmermann2015, Espinosa2016}.

In the seeding technique \cite{Bai2005, Bai2006a}, the critical nucleus
size \(N_0^*\) is determined at a particular temperature, using seeded
MD (Molecular Dynamics) simulations. Figure \ref{fig:seedingFlowChart}
outlines the procedure of the seeding technique. A solid spherical
cluster of the stable crystal phase (or `seed') is artificially inserted
into supercooled liquid configurations. The seed dimensions are chosen
prior to insertion. The evolution of cluster sizes of the largest
clusters, in a swarm of independent fleeting trajectories, is then
monitored at predetermined trial temperatures. The temperature at which
the `largest' clusters (determined using an appropriate order parameter)
grow in roughly half of the trajectories is the
temperature at which the inserted initial cluster size (\(N_0\) in
Figure \ref{fig:seedingFlowChart}) is critical ( that is, \(N_0=N_0^*\)
at \(T\)). This scenario corresponds to the top of the free energy
barrier, where a particular cluster of the ordered phase, with \(N_0^*\)
particles, may either re-disperse into the liquid phase or increase in
size. Therefore, using seeded simulations, the value of \(N_0^*\) is
fixed at a particular temperature \(T\).

The estimated \(N_0^*\) is subsequently used to obtain \(\sigma_0\),
assuming that the free-energy profile obeys CNT. Now the interfacial
energy \(\sigma_0\), can be determined according to the relation

\begin{equation}\label{eqn:chap3cntInterfacial}
  \sigma_0 = \left( \frac{3 N_0^*}{32 \pi v'^2} \right)^{\frac{1}{3}} |\Delta \mu_0|.
\end{equation}

The main advantage of the seeding technique is that it allows for a
direct `microscopic' estimate of the interfacial free energy for a wide
range of metastable conditions. In fact, the assumption that the nucleus
is spherical can be relaxed by calculating both the nucleus shape factor
and the critical nucleus size as free fitting parameters
\cite{Zimmermann2015, Zimmermann2018}. In another variant of the
standard approach to seeding, the committor probability, \(P_B (N_0)\),
can be fitted to obtain \(N_0^*\) at \(P_B = 0.5\) \cite{Richard2018}.

Although the seeding method improves the basic premise of CNT, it still
incorporates many of the characteristic weaknesses and assumptions of
CNT. Nevertheless, the seeding method has provided surprisingly accurate
results for a variety of diverse systems.


Figure \ref{fig:seedingWatNoShear} depicts the performance of the
standard seeding approach, as implemented by Espinosa et
al. \cite{Espinosa2014, Espinosa2016}, for the atomistic models
TIP4P/2005, TIP4P/Ice and the coarse-grained monatomic (mW) water model.
The percent supercooling for every model is defined as
\(\Delta T\% = \frac{T_m - T}{T_m} \times 100\) where \(T_m\) is the
melting point \cite{Goswami2021a}. Evidently, seeding method
calculations using the atomistic water models have yielded remarkably
good agreement with experimental data for water. Theoretical results
from FFS, and brute-force simulations agree with seeding calculations
within error bars, with one notable exception. The nucleation rate of
the TIP4P/Ice model, at a temperature of \(230 \ K\) obtained from
rigorous FFS calculations (filled turquoise diamond symbol in Figure
\ref{fig:seedingWatNoShear}) deviates significantly, by about \(11\)
orders of magnitude, from both experimental results and those from
seeding.

\section{\label{computational-strategies-for-sheared-nucleation}Computational Strategies for Sheared
Nucleation}

\subsection{\label{brute-force-approaches}Brute-Force Approaches}

Brute-force approaches constitute what is perhaps the most conceptually
straightforward way of obtaining critical nucleus information, and
steady-state nucleation rates. In essence, brute-force approaches
involve cooling down the system below the freezing point, and performing
isothermal constant-temperature simulations until a nucleation event is
observed. Some manipulation is required to extract information from an
ensemble of nucleation events---homogeneous nucleation is a stochastic
rare event by the standards of simulations---but it should be emphasized
that the dynamics of the system are not changed at all by the
application of brute-force techniques.

A common strategy for generating the requisite ensemble of nucleation
events is to perform several MD simulations from configurations at the
same initial conditions, monitoring the time evolution of these systems
until multiple nucleation events have been observed. Techniques differ
in how this collection of nucleation events is processed. It should be
noted that, from the methodological point of view, imposing shear on the
system does not affect the general procedure adopted in brute-force
approaches, although an ensemble of trajectories is required for each
shear rate (and the prevailing thermodynamic conditions). The only
concession required for the incorporation of shear is that sheared MD
simulations must now be performed in the \(NVT\) ensemble, using the
SLLOD algorithm \cite{Evans1984} and Lees-Edwards boundary conditions
\cite{Lees1972, Daivis2006}, instead of in the \(NPT\) ensemble, as is
typical for MD simulations of quiescent nucleating systems.

\subsubsection{\label{rates-from-the-survival-probability}Rates from the Survival
Probability}

In an approach that closely mirrors how rates are estimated from
quiescent nucleation experiments \cite{Brandel2015}, induction times are
calculated from an ensemble of nucleating trajectories. Induction times
can be deduced from the drop in the potential energy of the system
associated with the formation of the critical nucleus
\cite{Fitzner2015}. The induction times are then used to calculate the
survival probability, \(P_{sur}(t)\), which is the probability of
finding the system in the metastable state, such that it is completely
devoid of the solid phase \cite{Resnick2013}. If nucleation can be
interpreted as a Poisson process \cite{Resnick2013}, then the surviving
probability, \(P_{sur}(t)\) can be modelled by a stretched exponential
function in time, \(t\) \cite{Fitzner2015}:

\begin{equation}\label{eqn:surProbability}
  P_{sur}(t) = exp[-(t \overline{t_0})^{\beta'}],
\end{equation}

where $\overline{t_0}$ is the average induction time, and $\beta'$ is
a corrective parameter accounting for non-exponential kinetics. Both
$\overline{t_0}$ and \(\beta'\) can be gleaned from fits of the values
of \(P_{sur}(t)\) to Eq. \eqref{eqn:surProbability}. The nucleation
rate, \(J\), can subsequently be calculated from the relation
\cite{Resnick2013}:

\begin{equation}\label{eqn:rateSurvivalProbability}
  J = \frac{1}{\overline{t_0} V},
\end{equation}

where \(V\) is the volume of the liquid.

\subsubsection{\label{mean-first-passage-time-method}Mean First-Passage Time
Method}

\begin{figure}[H]
\centering
\includegraphics[width=0.6\textwidth]{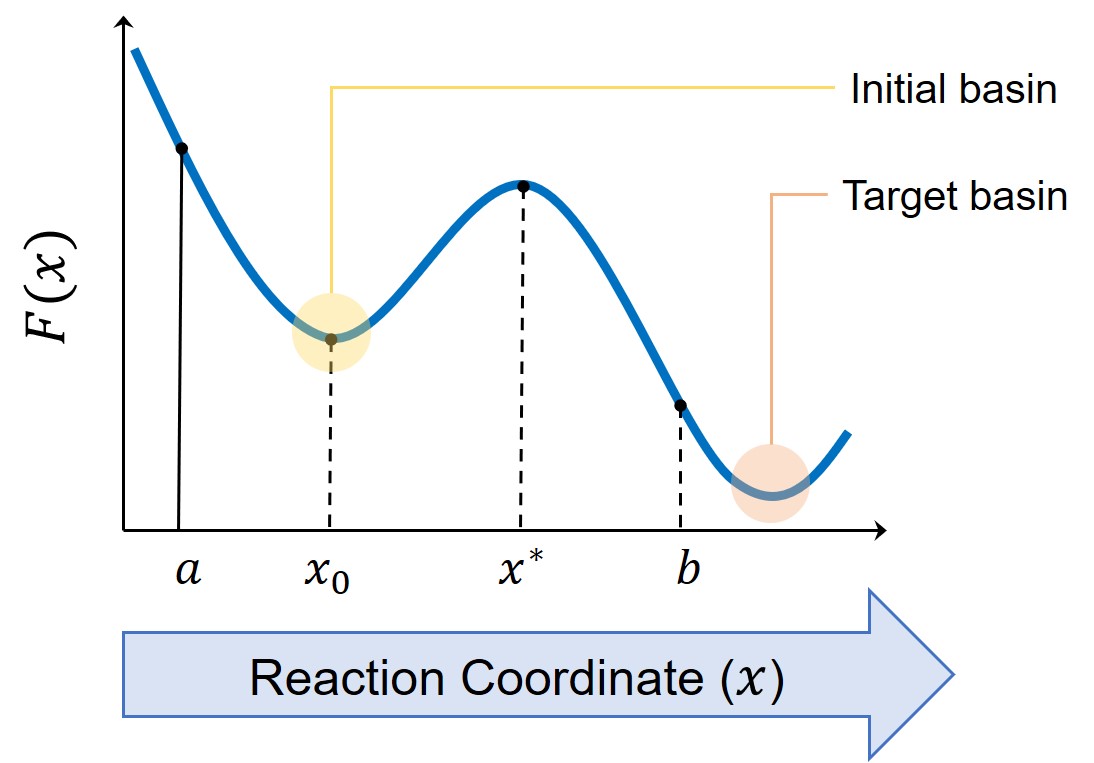}
\caption{Schematic representation of an activated process. The reflecting boundary is at $x=a$, and the absorbing boundary is at $x=b$. The transition state, $x^*$, is situated at the peak of the free energy barrier which separates the initial and target basins.}
\label{fig:mfptActivatedProcess}
\end{figure}

The crux of this method is the assumption that the nucleation kinetics is solely controlled by the height and curvature of a sufficiently large free energy barrier. Under these conditions, growth is much faster than nucleation, and the MFPT can be directly related to the steady-state rate of barrier crossing \cite{Wedekind2007}. In most treatments, the MFPT is denoted by \(\tau\), which we abstain from in
this review to avoid confusion with characteristic times associated with
sheared nucleation (Section
\ref{classical-nucleation-theory-approaches}). Here, we use the notation
\(\Tau\) for the MFPT. We represent the reaction coordinate, which maps
the progress of the system, by \(x\). In the jargon of Kramer's reaction
rate theory \cite{Haenggi1990}, the transition state, \(x^*\), is at the
top of the free energy barrier. The free energy barrier separates the
initial basin, the edge of which is a reflecting boundary at \(x=a\),
and the final target basin, which yields the absorbing boundary at
\(x=b\). The system is initially "trapped" in the initial basin, at a
starting point \(x_0\). This scenario is visually depicted by Figure
\ref{fig:mfptActivatedProcess}. The MFPT, $\Tau(x_0; a,b)$ is formally
defined as the average time taken by the system, initially starting from
\(x_0\), to leave a particular domain \([a,b]\) for the first time.

The premise of the MFPT formalism rests on certain preconditions. The
activation rates, or escape rates, from \(a\) are only properly defined
when the free energy barrier is sufficiently high (\(F(x^*)>>k_B T\),
where \(k_B T\) is the thermal energy). The direct consequence of this
condition is that there is a large degree of separation between the time
scale of the barrier-crossing event and that of thermal diffusion; in
other words, barrier-crossing is a stochastic thermally activated
process. This generally holds true for homogeneous nucleation. Within
this context, the MFPT, $\Tau(x)$, exhibits a distinctive sigmoidal
shape. The rate, \(J\), is very simply obtained from the inverse of
$\Tau(x^*)$, and is agnostic to both the initial condition \(x_0\) and
\(b\).

For both sheared and quiescent homogeneous nucleation, the reaction
coordinate usually chosen is the size of the largest solid cluster,
\(N\). Therefore, the analytical expression for the MFPT, $\Tau(N)$,
according to the MFPT formalism is given by \cite{Wedekind2007}:

\begin{equation}\label{eqn:mfpt}
  \Tau(N) = \frac{1}{2 J V} {1 + erf[c(N-N^*)]},
\end{equation}

where \(J\) is the nucleation rate, \(V\) is the total volume of the
system, \(N^*\) is the critical nucleus size for a particular condition
of metastability, and \(c=Z \sqrt{\pi}\).

A description of a general implementation of the MFPT method is
enumerated below:

\begin{enumerate}
\def\labelenumi{\arabic{enumi}.}
\item
  The first-passage time is determined for every nucleating run. The
  first-passage time is defined as the time \(t\) taken for a nucleus of
  a particular size to first appear in a trajectory. The largest solid
  cluster \(N\) is determined for every saved configuration in the
  nucleating run under consideration. The stochastic nature of embryo
  size evolution implies that the time series \(N(t)\) need not be
  monotonic, and tends to change by increments or decrements that are
  greater than unity. Skipped embryo sizes can be ``filled-in''
  \cite{Lundrigan2009}, so that if an embryo of a given size is observed
  at a certain time (say, \(t'\)), the same time \(t'\) is assigned to
  smaller sizes which have not yet manifested in the trajectory. This
  ``filling-in'' procedure ensures that every nucleating run generates a
  first-passage time such that \(N\) changes by increments of \(1\) and
  no embryo sizes are skipped. Averaging over the first-passage times
  for the entire ensemble of nucleating runs yields the mean
  first-passage time, $\Tau(n)$.
\item
  $\Tau(n)$, is subsequently fitted to the expression given by Eq.
  \eqref{eqn:mfpt}. \(J\), \(N^*\) and \(Z\) are obtained as fitting
  parameters.
\end{enumerate}

Luo et al. \cite{Luo2020} calculated the sheared homogeneous nucleation
rates of mW water, at a supercooling of \(67.6 \ K\), using both the
survival probability technique (Section
\ref{rates-from-the-survival-probability}) and MFPT (described above).
Disadvantages of the brute-force approaches are immediately obvious:
owing to the hefty computational cost of generating an ensemble of
nucleation events using unbiased MD simulations, either the
computational model or the condition of metastability must be
compromised. For instance, a coarse-grained model could be chosen over
more realistic models out of such practical considerations.

\subsection{\label{forward-flux-sampling}Forward-Flux Sampling}

\begin{figure}[H]
\centering
\includegraphics[width=0.9\textwidth]{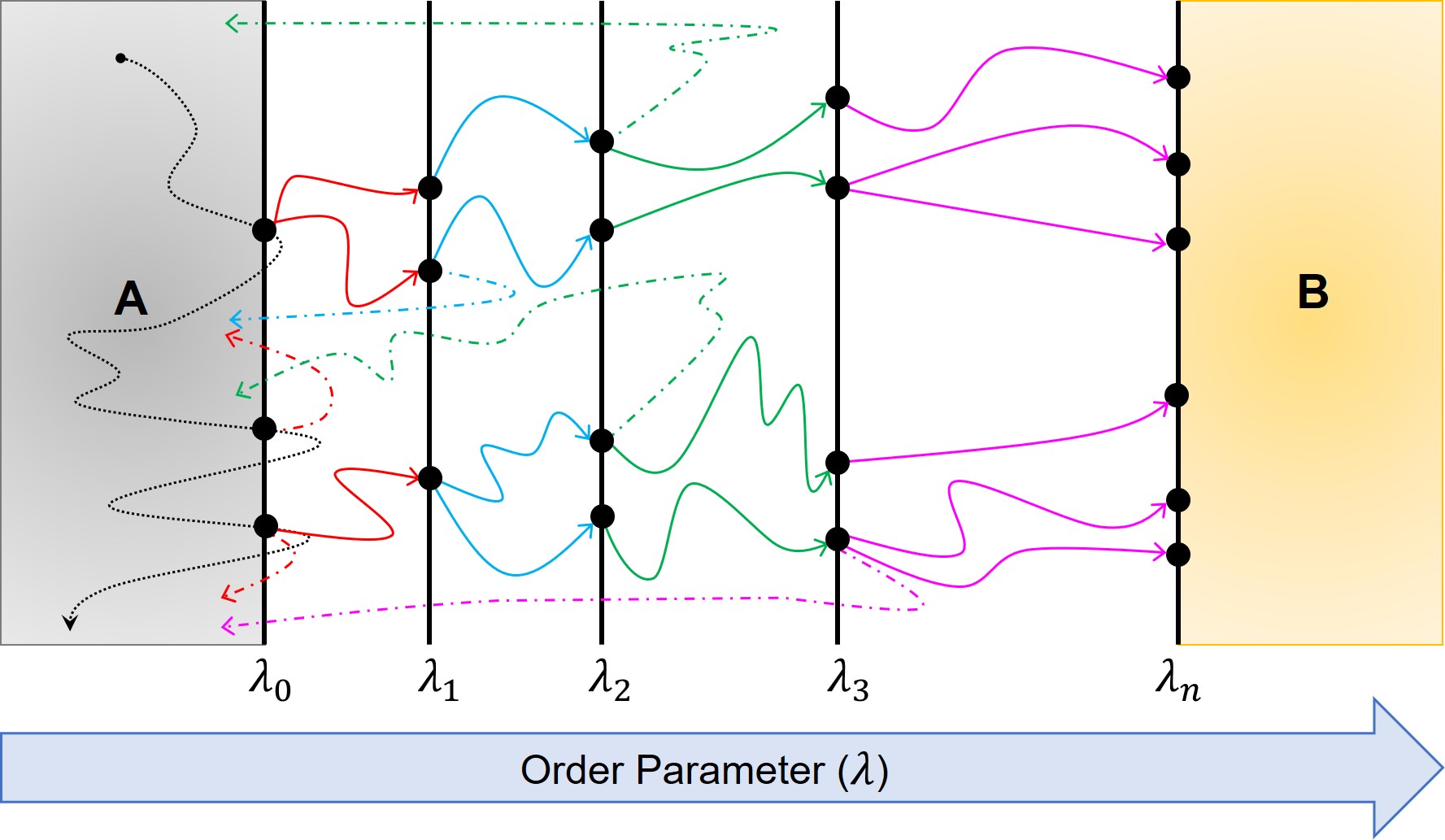}
\caption{Schematic illustration of the DFFS algorithm \cite{Allen2006}. The phase space between the initial state $A$ and the final state $B$ is partitioned by a series of $n$ non-overlapping interfaces ($\lambda_0, \lambda_1, \lambda_2...\lambda_n$), defined by an order parameter $\lambda$. The black dotted line corresponds to the initial MD trajectory, using which the escape flux, $\Phi$, is calculated. Solid and dashed lines denote the successful and unsuccessful runs, respectively. Each filled black circle represents a saved configuration on a particular interface. The different colours denote the order of trial runs, coloured according to the interface from which they are launched (in successive order red, blue, green, and magenta).}
\label{fig:ffs-schematic}
\end{figure}

Forward-flux sampling (FFS) \cite{Allen2005, Allen2006} is a robust
path-sampling approach which is capable of simulating rare events in
nonequilibrium systems with stochastic dynamics. Nonequilibrium systems
are characterized by the absence of detailed balance, which means that
they cannot be described by a Boltzmann-like stationary distribution and
do not satisfy the condition of time reversal symmetry. These features
render several rare event simulation techniques---for example,
Transition Path Sampling (TPS)
\cite{Dellago1998, Dellago1999, Bolhuis2002}, Transition Interface
Sampling (TIS) \cite{Erp2003, Erp2005}, Milestoning
\cite{Faradjian2004, West2007}--- unsuitable for nonequilibrium systems.
On the other hand, FFS can be used for both equilibrium and
nonequilibrium systems, provided that the underlying process is
Markovian. Naturally, this suggests that FFS could be an attractive
option for studying sheared homogeneous nucleation. FFS and its many
offshoots and flavours are covered in detail by extensive reviews
\cite{Allen2009, Hussain2020}. Here, we briefly cover the salient points of
the original FFS formulation or DFFS (Direct Forward Flux Sampling)
algorithm \cite{Allen2006}, pictorially depicted in Figure
\ref{fig:ffs-schematic}.

FFS provides a recipe for calculating the transition rate, \(k_{AB}\),
and for generating an ensemble of trajectories from an initial state
\(A\) to a final state \(B\). The transition rate, or "rate constant",
\(k_{AB}\), can be written in terms of two contributions: \cite{Erp2003}

\begin{equation}\label{eqn:transRateFFS}
  k_{AB} = \Phi P_B,
\end{equation}

where \(\Phi\) is the escape flux, defined as the flux of trajectories
which leave the initial state \(A\), and \(P_B\) is the 'transition
probability' or the probability that a trajectory departing from the
initial state \(A\) reaches the final state \(B\) without returning to
\(A\).

The initial and target basins \(A\) and \(B\) are defined with respect
to an appropriate order parameter, \(\lambda\), such that if
\(\lambda \le \lambda_A\) the system is in the initial state \(A\), and
if \(\lambda \ge \lambda_B\) the system lies in the final state \(B\).
The configuration space sandwiched by the initial and final states \(A\)
and \(B\) is defined by \(\lambda_A<\lambda<\lambda_B\).

In FFS, the configuration space between an initial state (\(A\)) and a
final state (\(B\)) is split by a sequence of non-intersecting
interfaces, defined by an order parameter, \(\lambda\). The initial
state, \(A\), is defined by \(\lambda < \lambda_A\), and the final
state, \(B\), by \(\lambda > \lambda_B\). However, the bottom line is
that the transition probability, \(P_B\), tends to be so small that
sampling it directly using conventional brute-force simulations is
prohibitive \cite{Kratzer2014}. Therefore, the phase space between the
basins \(A\) and \(B\) is split into a sequence of \(n\)
non-intersecting interfaces, \(\lambda_i\), such that successive
interfaces lie in the direction of increasing \(\lambda\) (that is,
\(\lambda_i<\lambda_{i+1}\)), and \(\lambda_0 \equiv \lambda_A\) and
\(\lambda_n \equiv \lambda_B\) (shown in Figure
\ref{fig:ffs-schematic}). Then \(P_B\) can be obtained from the
following relation:

\begin{equation}\label{eqn:transProbabilityFFS}
  P_B = \prod_{i=0}^{n-1} p_i,
\end{equation}

where \(p_i\) is the conditional probability that a trajectory launched
from interface \(i\) reaches the next interface \(i+1\) without
returning to the initial state \(A\). Clearly, it is much easier to
sample \(p_i\) compared to the much smaller \(P_B\). Thus, a branching
ensemble of transition paths is generated via DFFS. A crucial
fundamental difference between FFS and other similar path-sampling
techniques is that in FFS system dynamics are integrated forward in time
only, dispensing with the need for detailed balance. This in turn
enables FFS to be used for nonequilibrium systems.

The DFFS algorithm was employed to analyse the sheared nucleation of a
simple test case: the two-dimensional Ising model
\cite{Allen2008, Allen2008a}. Clearly, FFS is especially promising as a
technique for studying sheared homogeneous nucleation because it was
expressly formulated for nonequilibrium systems (while also applicable
to equilibrium systems), and exploits the fluctuations of the system
dynamics in the direction of the order parameter. However, a significant
drawback of FFS is its prohibitive cost \cite{Hussain2020}, which may make it an impractical choice for more complex systems under shear.

\subsection{\label{classical-nucleation-theory-approaches}Classical Nucleation Theory
Approaches}

Physical intuition suggests that a linear flow field enhances the
diffusion of molecules, while also injecting an additional elastic
energy into the system. This class of techniques adheres to the core
tenets of CNT, while incorporating modifications or extensions to
account for these effects of shear. In this section, we will primarily
expound recently proposed methods which quantify the elastic energy,
pinpointing the free energy cost of the applied shear as a reversible
elastic work term, \(W_S\). Differences between techniques arise from
how this elastic work \(W_S\), is calculated. In the interest of clarity, we will employ a consistent system of notations for each formalism discussed below.

\subsubsection{\label{rates-as-functions-of-the-shear-rate-in-the-cluster-radius-space}Rates as Functions of the Shear Rate in the Cluster Radius Space}

In the framework of Mura et al. \cite{Mura2016}, the free energy cost
arising from the application of a volume-preserving shear rate,
\(\dot{\gamma}\), on a solid nucleus is equated to the reversible
elastic work experienced by the liquid, \(W_L\). The applied shear rate
tends to deform the solid nucleus into an ellipsoid without changing the
volume contained by the nucleus. The free energy of a solid nucleus in a
bulk homogeneous system, subjected to a simple shear rate,
\(\dot{\gamma}\) is then given by \cite{Mura2016}:

\begin{equation}\label{eqn:freeEnergyRMura}
  F(R) = -\frac{4}{3} \pi R^3 \frac{|\Delta \mu_{0} |}{v'} + 4 \pi R^2 \beta \sigma_{0} + \underbrace{\frac{1}{2} G (\tau \dot{\gamma})^2 \frac{4}{3} \pi R^3}_{\textrm{elastic work}},
\end{equation}

where \(F(R)\) is the free energy of formation of a cluster of radius
\(R\), and the reversible elastic work is given by
\(W_L = \frac{1}{2} G (\tau \dot{\gamma})^2 \frac{4}{3} \pi R^3\).
\(\Delta \mu_{0}\) is the chemical potential difference between the
thermodynamically stable crystal phase and the metastable liquid phase
when no shear is applied, \(\sigma_{0}\) is the surface tension or the
interfacial free energy of the nucleus at zero shear, \(v'\) is the
volume of one molecule in the crystal phase, \(G\) is the shear modulus
of the nucleus. We define a characteristic time \(\tau\) \cite{Goswami2021a} as
\(\tau=\frac{\eta}{G}\), such that \(\eta\) is the fluid viscosity.
\(\beta = 1 + \frac{7}{24} (\tau \dot{\gamma})^2\) is a corrective
"shape factor" accounting for the shear-induced change in shape of the
nucleus from a sphere to an ellipsoid. For perfectly spherical nuclei,
\(\beta\) is unity \cite{Mura2016}.

The radius of the critical nucleus, \(R^*\), corresponding to the top of
the nucleation barrier, obtained from Eq. \eqref{eqn:freeEnergyRMura}
is:

\begin{equation}\label{eqn:criticalRMura}
  R^* = \frac{2 \beta \sigma_0}{\left[ \frac{|\Delta \mu_0|}{v'} - \frac{1}{2} G (\tau \dot{\gamma})^2 \right]}.
\end{equation}

The corresponding free energy barrier at the critical nucleus radius, \(R^*\), is
given by:

\begin{equation}\label{eqn:criticalFheightMura}
  F(R^*) = \frac{16 \pi \sigma_0^3}{3} \frac{\beta^3}{\left[ \frac{|\Delta \mu_0|}{v'} - \frac{1}{2} G (\tau \dot{\gamma})^2 \right]^2}.
\end{equation}

The steady-state nucleation rate, or current, \(J\), across the
nucleation barrier, can be determined using Kramer's escape rate theory
\cite{Kramers1940, Zwanzig2001}. The nucleation rate is found to be:

\begin{equation}\label{eqn:rateMura}
  J = N_{tot} \frac{\sigma_0}{k_B T} \left[ 8 + \frac{7}{3} (\tau \dot{\gamma})^2 \right] D(R^*) e^{-\frac{F(R^*)}{k_B T}},
\end{equation}

where \(N_{tot}\) is the total number of particles in the metastable
state, and \(D(R^*)\) is the diffusion coefficient for a critical
cluster of radius \(R^*\).

The kinetic prefactor is the quantity
\(\frac{\sigma_0}{k_B T} \left[ 8 + \frac{7}{3} (\tau \dot{\gamma})^2 \right] D(R^*)\)
in Eq. \eqref{eqn:criticalFheightMura}. The kinetic prefactor can be
estimated by a solution of the Smoluchowski equation in \(R\), in the
limit of small shear stresses. The calculation of the kinetic prefactor
requires quantifying the effective attraction driving the attachment of
particles to the crystal nucleus. Mura et al.~\cite{Mura2016} analysed a
model colloidal hard sphere system using this formalism. The authors
assumed that the effective attraction between a freely moving particle
in the liquid phase and a particle belonging to the solid nucleus is
completely described by the potential of mean force for hard spheres.
The attraction was modelled as a ramp potential.

The equations explicitly incorporate shear as a dependent variable, while
separating zero-shear quantities (\(\Delta \mu_0\), \(\sigma_0\) etc.)
from flow properties and mechanical properties (\(\eta\), \(G\), etc.).
Consequently, it is possible to analyse the effect of shear on the
nucleation rates without running simulations or performing calculations
specific to any particular shear rate. Therefore, the formalism of Mura
et al. \cite{Mura2016} provides an attractive and computationally
inexpensive alternative to brute-force approaches for studying the
crystallization of sheared liquids. However, a mechanism was not
provided within the framework for determining the interfacial energy,
\(\sigma_0\), which is generally a non-trivial calculation. In addition,
the estimation of the kinetic prefactor, and specifically of \(D(R^*)\),
can become more convoluted for more sophisticated atomic potentials.

\subsubsection{\label{size-dependent-shear-moduli-calculated-for-every-shear-rate}Size-Dependent Shear Moduli Calculated for Every Shear Rate}

Richard et al.~\cite{Richard2019} proposed another approach, tying
together seeded simulations performed at different shear rates with
modified CNT equations. In the framework of Richard et
al.~\cite{Richard2019}, the reversible elastic work of the solid phase,
\(W_S\), is not estimated from the shear stress acting on the
surrounding liquid phase, \(\nu_L=\eta \dot{\gamma}\). Instead, \(W_S\)
is calculated from simulations for every shear rate and condition of
metastability. It is assumed that the nuclei or `solid droplets' are
perfectly spherical. The reversible elastic work of a nucleus of radius
\(R\) and volume \(V_S = \frac{4}{3} R^3\) is given by:

\begin{equation}\label{eqn:elasticWorkRichard}
  W_S(V_S) = \frac{\nu_S^2}{2 G(V_S)} V_S,
\end{equation}

where \(G(V_S)\) is the volume-dependent shear modulus of the solid
droplet, and \(\nu_S\) is the shear stress experienced by the nucleus.
One of the salient features of this approach, which distinguishes it
from the formalism of Mura et al.~\cite{Mura2016}, is the assertion that
the shear modulus of finite clusters is not equivalent to the bulk solid
phase shear modulus. This is embodied by the calculation of an
`effective' shear modulus, \(G_{eff}\) for critical nuclei.

For small shear stresses, in the limit of the linear response regime,
the free energy barrier height is approximated by:

\begin{equation}\label{eqn:freeEnergyRichard}
  F(\nu_S) = F_0 (1 + a_W ),
\end{equation}

where \(a_W\) is a linear response coefficient, and \(F_0\) is the free
energy barrier in the absence of shear. \(a_W\) is dependent on liquid
and solid properties, predicted by CNT according to the expression

\begin{equation}\label{eqn:aWrelationRichard}
  a_W = \frac{1}{G_{eff} \Delta P},
\end{equation}

where \(\Delta P\) is the thermodynamic driving force for nucleation,
i.e., the pressure difference between the pressure inside the solid and
the ambient liquid pressure.

Using the CNT prediction of \(a_W\) given by Eq.
\eqref{eqn:aWrelationRichard}, the expression for the nucleation rate,
\(J\), can be obtained as:

\begin{equation}\label{eqn:rateRichard}
  lnJ \approx lnJ_0 + a_J \nu_S^2,
\end{equation}

where \(J_0\) is the nucleation rate at vanishing stress
\((\dot{\gamma}=0)\), and \(a_J\) is a linear response coefficient given
by \(a_J = -\Delta F_0 a_W\).

Once \(a_W\) is known, the linear response coefficient, \(a_J\), in Eq.
\eqref{eqn:rateRichard} can be determined. Therefore, the rate
calculations hinge on the estimation of the size-dependent shear
modulus, \(G_{eff}\).

For small solid droplets, the shear modulus is extracted using the
following relation:

\begin{equation}\label{eqn:GeffCalcRichard}
  G_{eff} = \frac{\nu_S}{\gamma_S},
\end{equation}

where \(\gamma_S\) is the solid strain experienced by the solid
droplets.

It is evident from Eq. \eqref{eqn:GeffCalcRichard} that the effective
shear modulus can be determined once the solid density, \(\rho_S\),
shear stress, \(\nu_S\), and solid strain, \(\gamma_S\), have been
estimated. Richard et al.~\cite{Richard2019} proposed a methodology for
extracting the required statistics via seeding simulations
\cite{Richard2018}, conducted at each condition of metastability,
imposed \(\dot{\gamma}\), and seed size considered.

Monodisperse hard spheres were studied using this formalism
\cite{Richard2019}. The authors validated the size-dependence of the
shear modulus by showing that the shear modulus of nuclei with sizes
less than \(100\) particles were significantly smaller than the shear
modulus of the bulk crystalline phase. For such small solid droplets,
the shear modulus (in reduced units) was found to be in the vicinity of
\(G \approx 40\), in comparison to the bulk shear modulus value of about
\(100\). However, a deviation from bulk properties is not surprising,
given the small size of the nuclei considered. Indeed, it is expected
for small nuclei to exhibit severe finite size effects. Deformation of
the solid nuclei was not considered, despite the fact that the surface
area contribution to the free energy is expected to be dominant for
small nuclei. Although the approach of Richard et al.~\cite{Richard2019}
is conceptually rigorous, accurate, and less computationally expensive
than brute-force methods, the framework may be limited in scope due to
the extensive calculations required for the estimation of \(G_{eff}\) at
every \(\dot{\gamma}\) considered.

\subsubsection{\label{rates-as-functions-of-the-shear-rate-in-the-cluster-size-space}Rates as Functions of the Shear Rate in the Cluster Size Space}

Goswami et al. \cite{Goswami2021a} proposed a formalism, similar in
spirit to that of Mura et al \cite{Mura2016}. Seeded simulations of the
quiescent metastable liquid form the backbone of this formulation,
wherein the shear rate appears as an explicit variable in the modified
CNT expressions. The methodology involves calculating the
shear-independent input quantities, along with transport properties
piece-by-piece to obtain the nucleation rates as functions of the shear
rate, \(\dot{\gamma}\).

This approach builds on the expression for the free energy cost,
\(F(R)\), described by Eq. \eqref{eqn:freeEnergyRMura}. The height of
the free energy barrier for sheared nucleation, corresponding to a
critical nucleus size \(N^*\) is given by \cite{Goswami2020a}:

\begin{equation}\label{eqn:criticalFGoswami}
  F(N^*) = \frac{N_0^* |\Delta \mu_0|}{2} \frac{\beta^3}{ \left[ 1- \frac{v'G}{1 |\Delta \mu_0|} (\tau \dot{\gamma})^2 \right] },
\end{equation}

where \(N_0^* = \frac{32 \pi \sigma_0^3 v'^2}{3 |\Delta \mu_0|^3}\) is
the critical nucleus size at zero shear, and
\(\beta = 1 + \frac{7}{24} (\tau \dot{\gamma})^2\) is the shape factor
correcting for ellipsoidal deformation.

The steady-state nucleation rate, \(J\), can be derived from
Zeldovich-Frenkel equation \cite{Kelton1983, Kelton1991}, yielding the
following familiar CNT-based expression \cite{Goswami2020a}:

\begin{equation}\label{eqn:rateGoswami}
  J = \rho_l Z f^+ e^{-\frac{F(N^*)}{k_B T}},
\end{equation}

where the nucleation rate \(J\) is the current or flux across the free
energy barrier, in the cluster-size space and is in units of the number
of nucleation events per unit volume per unit time, \(f^+\) is the rate
of attachment of particles to the critical cluster, \(\rho_l\) is the
number density of the supercooled liquid, \(Z\) is the Zeldovich factor,
and \((\rho_l Z f^+)\) is the kinetic prefactor. This rate equation is
an analogue of Eq. \eqref{eqn:chap3rateCNT}. However, here, \(Z\),
\(f^+\), \(N^*\) are functions of the shear rate, \(\dot{\gamma}\).

The attachment rate depends on the the two-dimensional diffusion
coefficient, \(D_l\), which in turn depends on the magnitude of
\(\dot{\gamma}\) at a particular temperature \(T\). It has been shown
that atomic systems and suspensions \cite{Siqueira2017, Chandran2020}
follow a linear relationship with \(\dot{\gamma}\), given by:

\begin{equation}\label{eqn:diffusionLinearGoswami}
  D_l = D_0 + c \dot{\gamma},
\end{equation}

where \(D_0\) is the diffusion coefficient at zero shear rate for the
quiescent liquid at a particular \(T\), and \(c\) is a fitting parameter
with units of squared length.

Using scaling arguments, the transcendental expression for \(J\) can be
approximated by a polynomial function in \(\dot{\gamma}\), valid for
shear rates for which Eq. holds true,
\(\dot{\gamma} < \frac{1}{\eta} \left( \frac{2G |\Delta \mu_0|}{v'} \right)^{\frac{1}{2}}\),
and \(\beta\approx1\) \cite{Goswami2021a}.

\begin{equation}\label{eqn:simpleRateGoswami}
  J = J_0 \left( 1 + \frac{c}{D_0} \dot{\gamma} \right) \left[1 - \frac{N_0^* v' G}{2k_B T} (\tau \dot{\gamma})^2 \right],
\end{equation}

where \(J_0\) is the nucleation rate when the shear rate is zero.

The form of Eq. \eqref{eqn:rateGoswami}, and more transparently, that of
Eq. \eqref{eqn:simpleRateGoswami} suggest that, at a particular
temperature, there exists a maximum in the nucleation rate at a specific
shear rate, \(\dot{\gamma_{opt}}\). An estimate of the dimensionless
optimal shear rate, \(\tau \dot{\gamma_{opt}}\) can be derived
analytically, for \(\frac{6 c^2 G k_B T}{N_0^* v' (D_0 \eta)^2}<<1\)
\cite{Goswami2021a}:

\begin{equation}\label{eqn:simpleGammaTauGoswami}
  \tau \dot{\gamma}_{opt} = \left( \frac{k_B T}{D_0 \eta} \frac{c}{v'} \right) \times \frac{1}{N_0^*},
\end{equation}

where the dimensionless group defined as
\(B = \left( \frac{k_B T}{D_0 \eta} \frac{c}{v'} \right)\) is related to
the transport properties (\(\eta\) and \(D_0\)). \(N_0^*\) is dependent
on the thermodynamic properties (\(\Delta \mu_0\), \(\sigma_0\) etc.).
The dimensionless product, \(\tau \dot{\gamma_{opt}}\), is an intrinsic
measure of how the temperature affects the variation of the nucleation
rates with \(\dot{\gamma}\).

A crucial difference between the approach of Goswami et al. \cite{Goswami2020a, Goswami2021a} and that of
Richard et al. \cite{Richard2019} is that seeded simulations must be
performed for every shear rate in the case of the latter, while for the
former only seeded simulations at zero shear are required. In addition,
the crux of the methodology of Richard et al.~\cite{Richard2019} is that
the size dependent shear modulus, \(G_{eff}\), is appreciably divergent
from the bulk crystal shear modulus.

Interestingly, Goswami et al.~\cite{Goswami2021a} found that predictions
from their formalism agreed very well, within error bars, with
brute-force MFPT calculations for the mW water model at \(207 \ K\)
\cite{Luo2020} (corresponding to a supercooling of \(67.6 \ K\)).

\begin{figure}[H]
\centering
\includegraphics[width=0.5\textwidth]{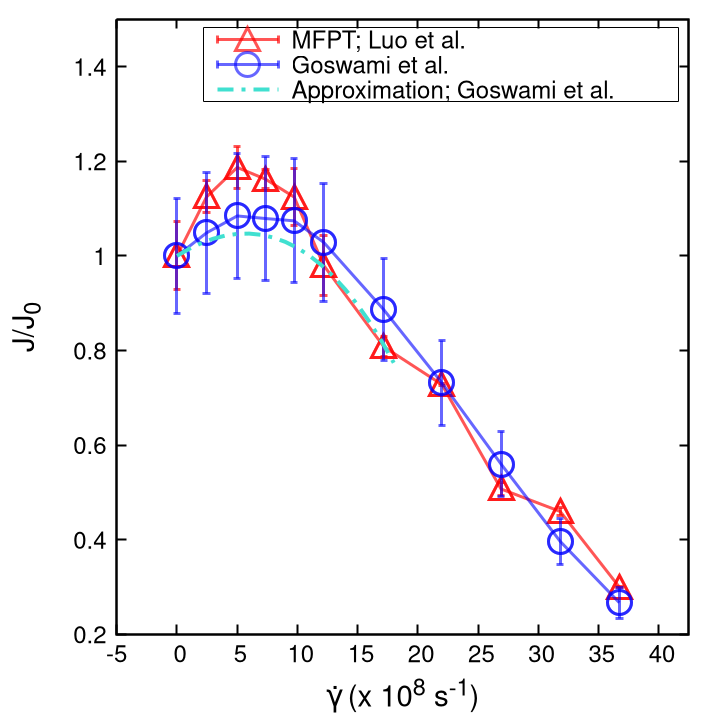}
\caption{Comparison of $J/J_0$ for the mW model at a supercooling of $67.6 \ K$, estimated using the formalism of Goswami et al. \cite{Goswami2020a, Goswami2021a} (open blue circles) with the MFPT results (open red triangles) of Luo et al \cite{Luo2020}. The polynomial approximation (Eq. \eqref{eqn:simpleRateGoswami}), depicted as a turquoise dashed line, agrees well with both sets of results.}
\label{fig:compareJbyJ0goswami}
\end{figure}

Figure \ref{fig:compareJbyJ0goswami} shows a comparison of the scaled
nucleation rates, \(J/J_0\), obtained using MFPT by Luo et al.~and those
calculated using Eq. \eqref{eqn:rateGoswami} and Eq.
\eqref{eqn:simpleRateGoswami}. Even the simple approximation described
by Eq. \eqref{eqn:simpleRateGoswami} shows high fidelity with
calculations performed using Eq. \eqref{eqn:rateGoswami}, within the
range of applicability.

While the magnitude of \(J_0\) depends solely on thermodynamic
quantities, including \(\Delta \mu_0\) and \(\sigma_0\), the shear
dependent \(J\) is influenced by shear-dependent quantities, and the
ratio \(J/J_0\) reveals the effects of shear on the nucleation rates.
Finite size effects, due to the size-dependent shear modulus and density
within the nuclei would presumably affect the variation of \(J/J_0\).
However, the good agreement with MFPT calculations shown in Figure
\ref{fig:compareJbyJ0goswami} suggests that the disparity in the bulk
modulus of finite clusters, reported by Richard et
al.~\cite{Richard2019} for hard spheres, does not significantly detract
from the predictive power of the formalism of Goswami et
al.~\cite{Goswami2021a}, at least for mW water, considering \(N_0^*\)
values of about \(70\). Finite-size effects are expected to be even less
significant for larger \(N_0^*\), corresponding to more realistic
supercoolings.

In an earlier work on colloidal suspensions
\cite{Blaak2004, Blaak2004a}, results from equilibrium umbrella sampling
calculations and non-equilibrium Brownian simulations were fitted
empirically to functions of \(\dot{\gamma}\). Shear-dependent
\(\Delta \mu\) and \(\sigma\) were obtained, which were subsequently
substituted in the original CNT equations (Section
\ref{overview-of-conventional-cnt}). The expressions for the `effective'
chemical potential and interfacial energy are given by:

\begin{equation}\label{eqn:chemPotBlaak}
  |\Delta \mu| = |\Delta \mu_0| (1 - c_0 \dot{\gamma}^2),
\end{equation}

\begin{equation}\label{eqn:intEnergyBlaak}
  \sigma = \sigma_0 (1 + \kappa_0 \dot{\gamma}^2),
\end{equation}

where \(c_0\) and \(\kappa_0\) are positive coefficients.

The governing equations of both Mura et al.\cite{Mura2016} and Goswami
et al.\cite{Goswami2020a} are consistent with this description:
quadratic functions of \(\dot{\gamma}\) can be defined for the effective
\(\Delta \mu\) and \(\sigma\), in terms of \(\Delta \mu_0\) and
\(\sigma_0\), respectively. Accordingly, the values of the coefficients
from both these approaches are \(c_0 = \frac{1}{2} G v' \tau^2\) and
\(\kappa_0= \frac{7}{24} \tau^2\).

\section{\label{results-from-simulations-and-theory}Results from Simulations and Theory}

\subsection{\label{effects-of-shear-on-energetics-kinetics-and-nucleation-rates}Effects of Shear on Energetics, Kinetics and Nucleation Rates}

\begin{figure}[H]
\centering
\includegraphics[width=\textwidth]{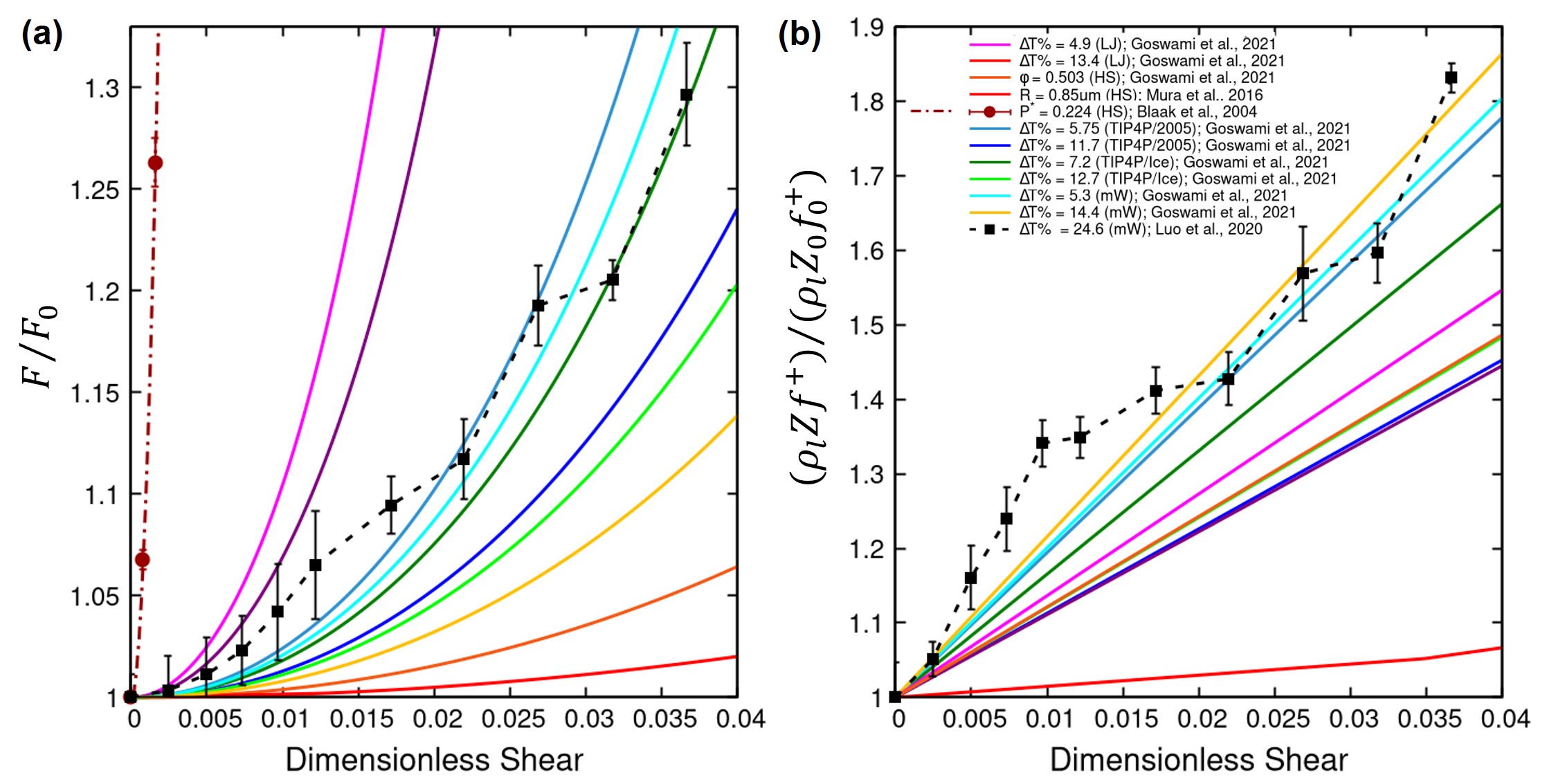}
\caption{(a) Increase of the normalized free energy barrier height, $F/F_0$, with shear. (b) Variation of the normalized kinetic prefactor, $(\rho_l Z f^+)/(\rho_l Z_0 f_0^+)$, with respect to the dimensionless shear rate. Trends for LJ, HS, colloids, atomistic water models (TIP4P/2005 and TIP4P/Ice), and the coarse-grained mW water model have been shown at several metastable conditions, using the results of Blaak et al. \cite{Blaak2004, Blaak2004a}, Mura et al. \cite{Luo2020}, Luo et al \cite{Luo2020}. and Goswami et al \cite{Goswami2021a}. Here, $\Delta T\%$, $P^*$, $R$ denote the percent supercooling \cite{Goswami2021a}, the pressure in reduced units for colloids \cite{Blaak2004, Blaak2004a}, and the radius of the hard spheres considered \cite{Mura2016}, respectively. The shear rates are multiplied by $\tau=\eta/G$ to obtain the dimensionless shear unless nondimensionalized shear rates are reported \cite{Mura2016, Luo2020, Goswami2021a}. For the mW model, at a percent supercooling of $24.6\%$ (filled black squares), $\eta = 30.94 \ mPas$ \cite{Goswami2021a} and $G = 3.1 \ GPa$ \cite{Cao2018} have been used.}
\label{fig:opposingEffectShear}
\end{figure}

From a molecular standpoint, a steady flow has complex energetic and
kinetic effects on the process of sheared homogeneous nucleation.
CNT-based approaches and MFPT conveniently split these shear
contributions into distinct trends affecting two separate quantities,
which can be analyzed independently: the free energy barrier and the
kinetic prefactor.

Figure \ref{fig:opposingEffectShear}(a) depicts the variation of the
free energy barrier with shear, drawing from various results on Lennard-Jones (LJ), hard spheres (HS),
colloids, and water models
\cite{Blaak2004, Blaak2004, Mura2016, Luo2020,Goswami2021a}. Clearly,
the free energy barrier height increases with increasing shear for every
system considered, irrespective of the specific methodology employed.
Blaak et al \cite{Blaak2004, Blaak2004a}. showed that the free energy
barrier is a quadratically increasing function of shear for colloids
modelled with Yukawa repulsion, which is a trend that is qualitatively
supported by subsequent studies on several other systems (Figure
\ref{fig:opposingEffectShear}(a)). Further, we note that the increase in
the free energy is also heralded by the dependence on \(\dot{\gamma}\)
in Eq. \eqref{eqn:criticalFheightMura} and Eq.
\eqref{eqn:criticalFGoswami}. A physical interpretation of this scenario
is that the imposed shear induces a small elastic deformation in the
solid nucleus, which incurs an additional free energy cost, thereby
enlarging the free energy barrier height (compared to the quiescent
nucleation free energy). The concomitant increase in the critical
nucleus with shear can also be rationalized by the shear deformation of
the nucleus
\cite{Blaak2004, Blaak2004, Mura2016, Luo2020, Goswami2020a, Goswami2021a, Yang2021}.

On the other hand, the kinetic prefactor is the embodiment of the
kinetics of the sheared nucleation process. Figure
\ref{fig:opposingEffectShear}(b) shows the increase of the kinetic
prefactor with increasing shear, which is a natural consequence of the
shear-enhanced motions of particles.

We surmise that the height of the free energy barrier and the kinetic
prefactor both tend to increase continuously with increasing
\(\dot{\gamma}\), at a specific condition of metastability (for example,
the prevailing \(T\), \(P\) or \(\phi\)). These trends indicate the
existence of a maximum in the nucleation rate, \(J\), at a particular
value of \(\dot{\gamma}_{opt}\). For shear rates such that
\(\dot{\gamma}<\dot{\gamma}_{opt}\), the kinetic enhancement dominates
the extra energetic cost arising from elastic deformation of the solid
nucleus, and \(J\) increases. \(J\) reaches its peak value at the
optimal shear rate, \(\dot{\gamma}_{opt}\) (at the thermodynamic
metastability condition \(T\), \(P\) or \(\phi\)). However, when the
applied \(\dot{\gamma}>\dot{\gamma}_{opt}\), the energetic effects of
shear overpower the kinetic increase, and \(J\) decreases. This
mechanism is corroborated by the non-monotonic behaviour of \(J\) with
\(\dot{\gamma}\), independent of the calculation method employed,
reported for the sheared two-dimensional Ising model \cite{Allen2008},
colloidal models \cite{Cerda2008, Lander2013}, hard spheres
\cite{Richard2015, Mura2016, Goswami2021a}, glassy systems
\cite{Mokshin2010, Mokshin2013}, a binary-alloy \cite{Peng2017}, more
recently mW water under shear
\cite{Luo2020, Goswami2020a, Goswami2021a}, and LJ and atomistic water
models \cite{Goswami2021a}.

\subsection{\label{temperature-dependence}Temperature Dependence}

\begin{figure}[H]
\centering
\includegraphics[width=\textwidth]{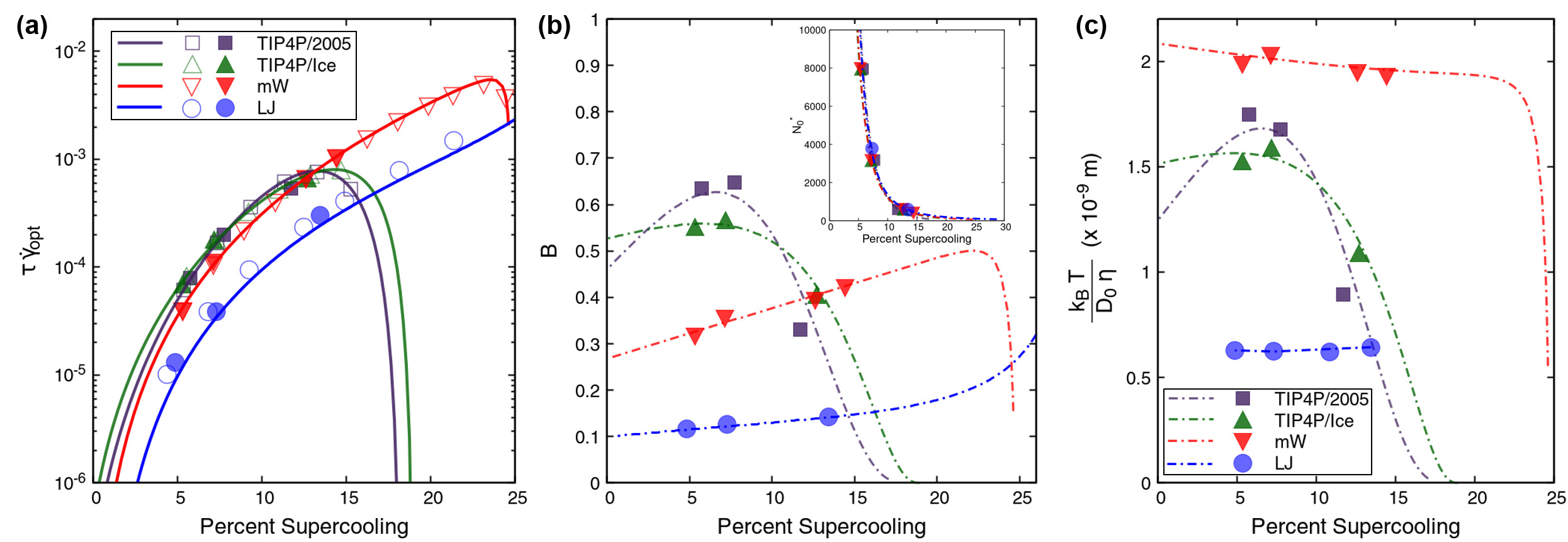}
\caption{(a) Variation of the dimensionless $\tau \dot{\gamma}_{opt}$ with the  percent supercooling, $\Delta T\%$, for atomistic water models, the mW water model and LJ. (b) Change in the dimensionless quantity, $B = \left( \frac{k_B T}{D_0 \eta} \frac{c}{v'} \right)$, with $
\Delta T\%$. The inset depicts the more-or-less 'universal' trend of $N_0^*$ with $\Delta T\%$. (c) Evaluation of the validity of the Stokes-Einstein (SE) relation, which states that $\frac{k_B T}{D_0 \eta}$ is constant with respect to temperature. The SE relation is violated for the water models, at deep supercoolings. Filled markers, open symbols and dashed lines denote results from simulation data, calculations using fitted equations to simulation data, and estimations from the approximate expression of Eq. \eqref{eqn:simpleGammaTauGoswami}, respectively. Reproduced with permission from Goswami et al \cite{Goswami2021a}. Copyright 2021 American Physical Society.}
\label{fig:gammaTauShearPRL}
\end{figure}

The dependence of the nucleation rate for quiescent systems, \(J_0\), on
\(T\) can be ascertained by using fits to the CNT expression (Eq.
\eqref{eqn:chap3rateCNT}) \cite{Espinosa2014, Espinosa2016}. The trend
of \(J_0\) with \(T\) for various water models is shown in Figure
\ref{fig:seedingWatNoShear}.

However, the homogeneous sheared nucleation rate, \(J\), is controlled
by several shear and temperature dependent parameters. To examine the
dual influences of temperature and shear, the dimensionless product,
\(\tau \dot{\gamma}_{opt}\) (given by Eq.
\eqref{eqn:simpleGammaTauGoswami}) was analyzed for water and LJ, using
a CNT-based approach (described in Section
\ref{rates-as-functions-of-the-shear-rate-in-the-cluster-size-space})
\cite{Goswami2021a}. The dimensionless optimal shear,
\(\tau \dot{\gamma}_{opt}\), can be considered to be a measure of the
relative `shift' of scaled nucleation rate curves (each at a particular
temperature), with shear. The percent supercooling, \(\Delta T\%\),
provides a convenient system-agnostic measure of the supercooling
\cite{Goswami2021a}.

Figure \ref{fig:gammaTauShearPRL}(a) shows the variation of
\(\tau \dot{\gamma}_{opt}\) with \(\Delta T\%\) for the TIP4P/2005,
TIP4P/Ice, mW and LJ models. It is strikingly apparent that the water
models exhibit different behaviour from that of LJ:
\(\tau \dot{\gamma}_{opt}\) is non-monotonic with \(\Delta T\%\) for
water. This translates to the non-monotonic behaviour in the optimal
shear rate, \(\dot{\gamma}_{opt}\), with temperature, for each water
model. This non-monotonicity was purported to be a new anomaly of water
\cite{Goswami2021a}. To investigate the cause of this anomalous trend,
the two constituent groups, \(B\) and \(N_0^*\), in Eq.
\eqref{eqn:simpleGammaTauGoswami} were analyzed. It was found that \(B\)
reflected the observed trends of \(\tau \dot{\gamma}_{opt}\), while the
variation of \(N_0^*\) with \(\Delta T\%\) was almost identical for both
water and LJ, as depicted in Figure \ref{fig:gammaTauShearPRL}(b).

Water is notoriously peculiar compared to most simple liquids,
exhibiting several anomalies in the supercooled regime
\cite{Pettersson2016}. One of the most well-known anomalies in the
dynamics of supercooled water is the breakdown of the Stokes-Einstein
(SE) relation. The SE relation asserts that \(\eta\) and \(D_0\) are
coupled such that \(\frac{k_B T}{D_0 \eta}\) is constant with
temperature \cite{Sutherland1905, Hynes1977}. Interestingly, the origin
of the non-monotonicity in \(\tau \dot{\gamma}_{opt}\) was directly
traced to the violation of the SE relation \cite{Goswami2021a}, shown in
Figure \ref{fig:gammaTauShearPRL}(c). The results suggest that, in
general, any system violating the SE relation (including glass-forming
liquids, and simple liquids near the glass transition) will exhibit the
same anomaly, that is, non-monotonicity of \(\tau \dot{\gamma}_{opt}\) with
temperature.

\subsection{\label{dependence-of-supersaturation}Dependence of
Supersaturation}

Colloidal suspensions are often modelled by hard sphere particles. For
HS, the packing fraction \(\phi\), indicating the degree of
supersaturation, is the condition of metastability dictating quiescent
nucleation. Therefore, \(\phi\) is analogous to the conditions of
metastability---\(T\) or \(P\)---for atomic systems \cite{Mura2016}. The
dependence of the rich quiescent nucleation behaviour of HS on \(\phi\)
has been well-documented in the literature. Phase transitions from the
disordered fluid state to crystal state take place in the packing
fraction range \(0.492 <\phi\le0.545\)
\cite{Alder1957, Alder1968, Gasser2001, Dullens2006, Vega2007}. Two
regimes for quiescent nucleation were discovered, for low and high
supersaturations
\cite{Pusey2009, Zaccarelli2009, Valeriani2011, Valeriani2012}. At lower
concentrations (\(\phi<0.56\)), standard nucleation and growth were
observed. At higher concentrations (\(\phi>0.56\)) a spinodal-like
nucleation regime was determined, wherein the free energy barrier is
virtually negligible.

Interestingly, the same crossover packing fraction, \(\phi=0.56\),
resolves the sheared homogeneous nucleation behaviour of HS into two
separate regimes. Richard et al \cite{Richard2015} examined the sheared
homogeneous nucleation of monodisperse HS, for various \(\phi\). The
chosen shear rates (\(0<\dot{\gamma}\le1.414\), where \(\dot{\gamma}\)
is in reduced units) were smaller than those required for shear-induced
ordering \cite{Vermant2005}. Two distinct regimes were identified,
corresponding to low and high supersaturation: 1) suppression of
crystallization, for \(\phi\le0.56\), and 2) crystallization
enhancement, in the range of \(0.56<\phi\le0.587\). The results clearly
indicate that the supersaturation plays a pivotal role in the sheared
homogeneous nucleation of HS. It is likely that supersaturation and
polydispersity are crucial factors governing the sheared nucleation of
hard-sphere colloids and colloidal glasses.

\subsection{\label{structural-effects-of-shear}Structural Effects of
Shear}

The structural consequences of shear on solid nuclei can be broadly
delineated into two related effects: changes in the overall shape, and
shear-induced changes in the microscopic structure of incipient nuclei.

\subsubsection{\label{shear-induced-changes-in-nucleus-shape}Shear-Induced Changes in Nucleus
Shape}

\begin{figure}[H]
\centering
\includegraphics[width=0.9\textwidth]{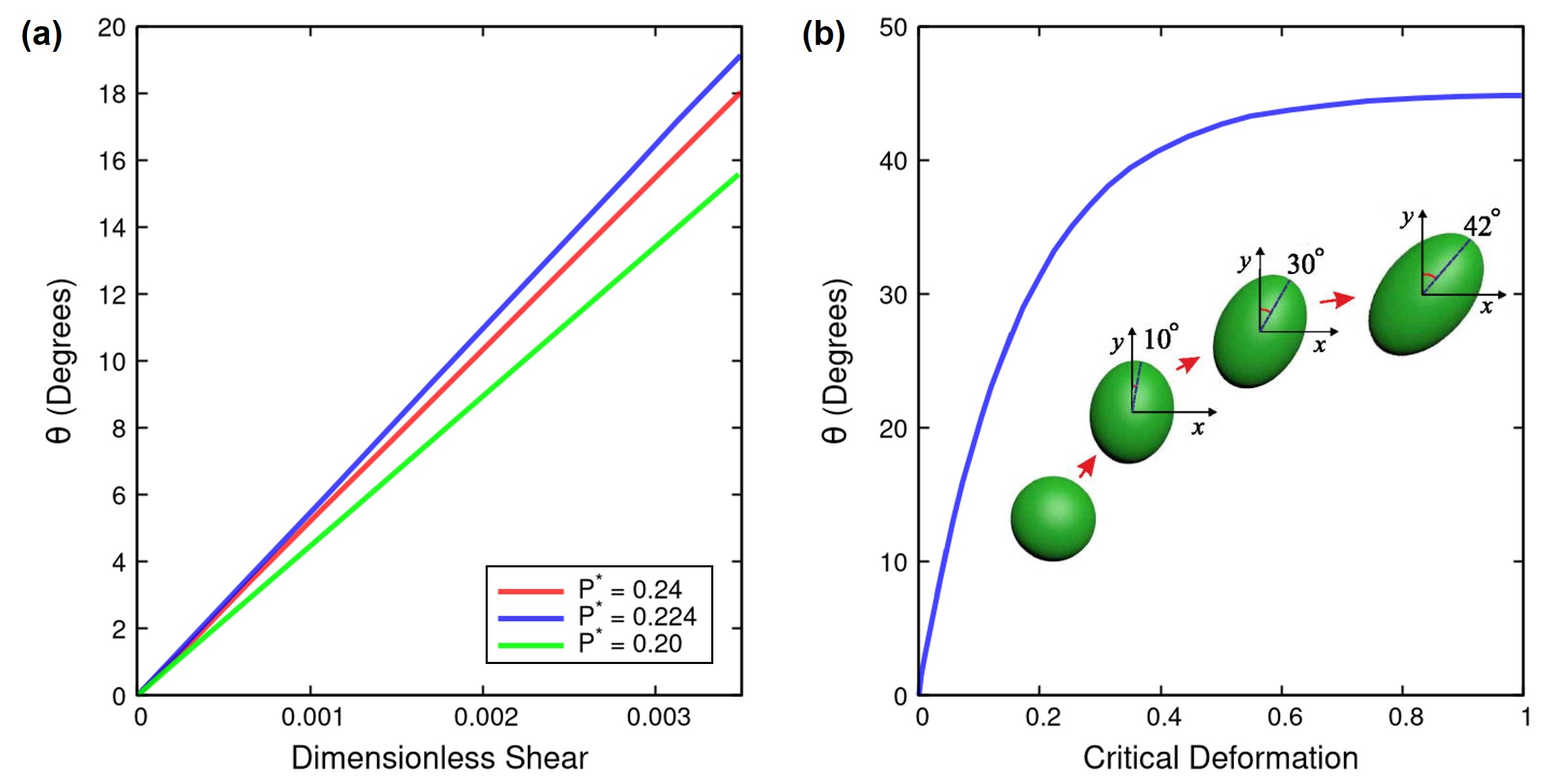}
\caption{(a) The tilt angle, $\theta$, plotted against the dimensionless shear for a colloidal system at various pressures (in reduced units), taken from Blaak et al \cite{Blaak2004, Blaak2004a}. The shear flow is in the $x$-dimension. (b) Power law fits of the tilt angle, $\theta$, to data for a model glassy system, at various temperatures, as a function of the critical deformation \cite{Galimzyanov2018}. The critical deformation is defined as the deformation of a critically sized nucleus, which is a dimensionless quantity. The insets depict visualizations of the change in the  orientations and shape of the critical nuclei. The direction of the applied shear, and the shear gradient are the $x$ and $y$ directions, respectively. Reproduced from Galimzyanov et al. \cite{Galimzyanov2018}, with the permission of AIP Publishing.}
\label{fig:tiltAngleLit}
\end{figure}

Changes in the shape of solid nuclei due to an imposed flow field have
been hypothesized, modelled and reported in theory, simulations, and
experiments for diverse systems, notably including polymers
\cite{BINSBERGEN1966, Kimata2007, Mykhaylyk2008, Read2020}.

The formation of elongated nuclei was observed in Brownian dynamics
simulations of colloids, which were deformed such that their largest
dimension was preferentially aligned along the direction of vorticity
\cite{Blaak2004, Blaak2004a}. Kinetic Monte-Carlo simulations of polymer
melts under shear revealed the growth of non-spherical nuclei in the
so-called "shish-kebab" morphology
\cite{BINSBERGEN1966, Hsiao2005, Kimata2007, Mykhaylyk2008, Graham2009, Dargazany2014}. On the other hand, the shape factor, \(\beta\),
(Section
\ref{rates-as-functions-of-the-shear-rate-in-the-cluster-radius-space})
was designed to theoretically account for the ellipsoidal deformation of
spherical nuclei in sheared simple liquids. Ellipsoidal shape
deformations were also found to occur for single-component glassy
systems \cite{Galimzyanov2018}.

The tilt angle, \(\theta\), is defined as the angle between the
shear-gradient direction and the longest axis of a solid ellipsoidal
nucleus \cite{Blaak2004}. Figure \ref{fig:tiltAngleLit}(a) and Figure
\ref{fig:tiltAngleLit}(b) represent how the shapes of nuclei are altered
due to a simple linear flow applied in the \(x\) direction, for colloids
\cite{Blaak2004, Blaak2004a} and a glassy system \cite{Galimzyanov2018},
respectively. Unsurprisingly, the shape deformations are more pronounced
for higher shear rates for both systems. We conclude that anisotropy in
the shape of nuclei is expected to be more prominent at high shear
rates. It stands to reason that computational approaches which assume
perfectly spherical nuclei under the action of shear should be applied
in the regimes of 'low' shear rates where these shape deformations are
less significant.

\subsubsection{\label{shear-induced-changes-in-microscopic-structure}Shear-Induced Changes in Microscopic
Structure}

The application of a simple shear rate, \(\dot{\gamma}\), imparts an
elastic deformation to a solid nucleus. It is not a stretch of logic to
expect that a mechanical deformation would also affect the internal
microscopic structure of emerging solid nuclei. The analysis of
microscopic structures is impervious to CNT-based approaches which, in
fact, usually neglect structural effects completely in rate
calculations. However, stacking faults and fivefold defect structures
\cite{Koning2020}, detected during the quiescent nucleation of
supercooled water \cite{Pingua2019, Johnston2012, Goswami2021}, have
been found to emerge as precursors to nucleation, even before critical
nuclei appear \cite{Li2011}. Furthermore, there is long-standing
evidence that morphological and structural characteristics of emergent
crystalline structures have a non-trivial impact on the sheared
nucleation process \cite{Kumaraswamy2002}. Theoretical inferences
notwithstanding, the possibility of shear affecting nascent crystal
defect structures, thereby influencing the nucleation pathway, has not
yet been explored for sheared homogeneous liquid-to-solid phase
transitions of simple liquids.

In this review, we have mainly focussed on the effect of a constant
simple shear rate on liquid-to-solid homogeneous phase transitions.
However, amorphous or disordered solid states can also undergo
solid-to-solid phase transitions. Deeply supercooled glassy systems may
be heated up (vitrification) or subjected to mechanical stress or shock
to aid the internal structural reshuffling required for a solid phase
transformation \cite{Schmelzer2005}. Shear-induced amorphous crystal
phase transformations have been investigated using experiments
\cite{Shao2015} and brute-force simulation approaches, for
single-component LJ systems \cite{Duff2007, Mokshin2008}, a
model metallic glass \cite{Mokshin2009}, and relatively stable
disordered systems like amorphous silicon (\(a-\textrm{Si}\))
\cite{Kerrache2011}. Such solid-to-solid phase transitions are
qualitatively distinct from phase transitions from the liquid state, and
exhibit intriguingly intricate phase behaviour. Employing molecular
dynamics simulations on a model of \(a-\textrm{Si}\), Kerrache et al
\cite{Kerrache2011}. identified three different regimes of shear-induced
structural reorganization: (i) formation of a more disordered state,
rich in fivefold liquid-like atoms at low temperatures or high shear
rates, (ii) the formation of an annealed state when thermal effects
dominate the effects of shear deformations, and (iii) the creation of a
crystalline solid phase. It is evident that subtle structural changes
dictate the nature of these solid-to-solid phase transitions, although
perhaps a path-sampling technique like FFS (alluded to in Section
\ref{forward-flux-sampling}) might be capable of determining mechanisms
of the proposed pathways.

\subsection{\label{polymer-crystallization-under-shear}Polymer Crystallization under
Shear}

\begin{figure}[H]
\centering
\includegraphics[width=0.6\textwidth]{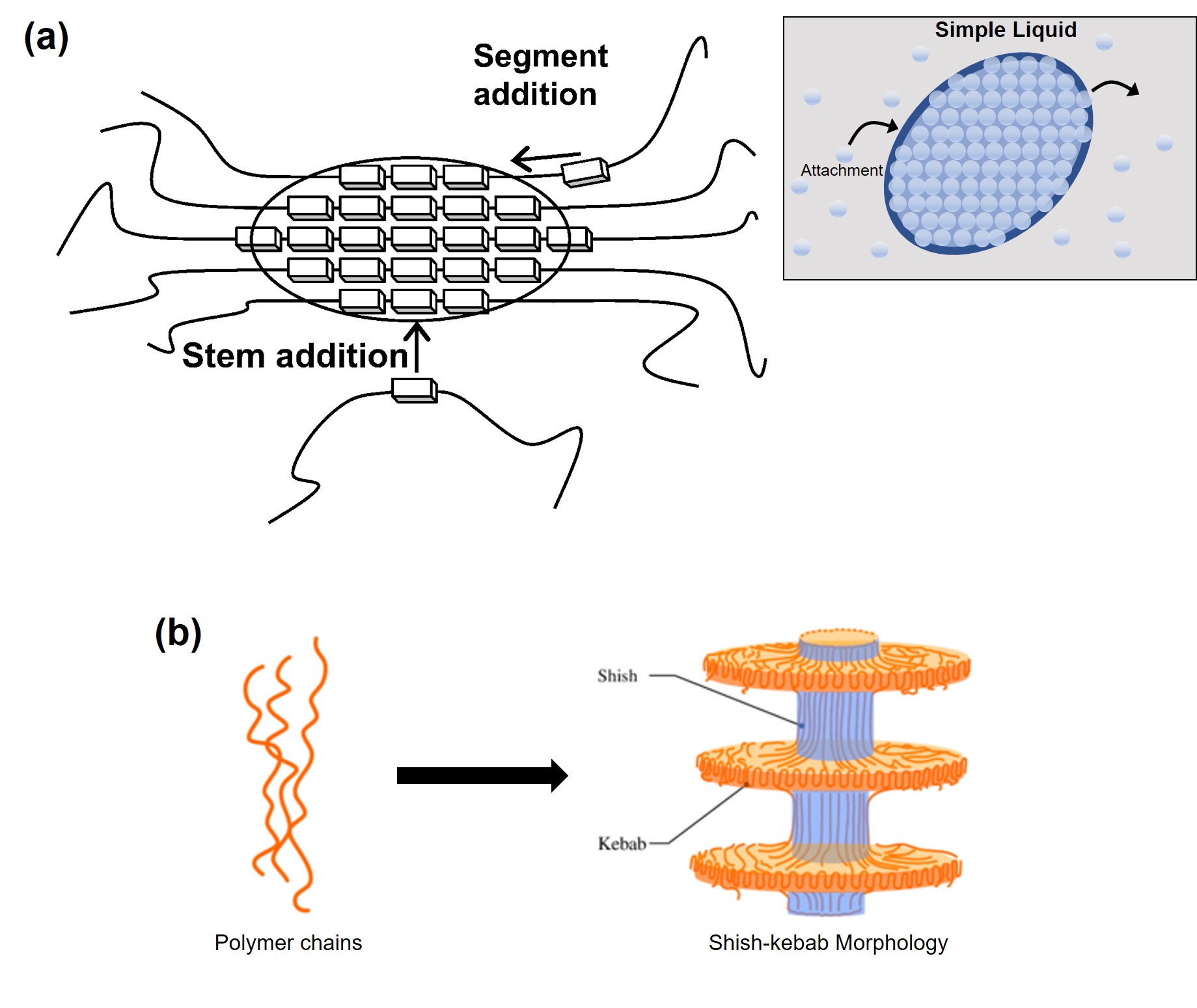}
\caption{(a) Schematic of the mechanism of polymer nucleus growth, as modelled by the coarse-grained GO model \cite{Graham2009, Graham2010a}. Reproduced from Ref \cite{Graham2014} by permission of The Royal Society of Chemistry. The inset depicts the attachment (and detachment) of molecules or particles, schematically represented by circles, to a shear-deformed ellipsoidal nucleus, according to CNT-based approaches \cite{Mura2016, Richard2019, Goswami2021a}. (b) Pictorial representation of the non-spherical "shish-kebab" morphology characteristic of sheared nucleation in polymers. Adapted with permission from Ref \cite{Dargazany2014}. Copyrighted by the American Physical Society.}
\label{fig:polymerShape}
\end{figure}

This review has concentrated mostly on how shear affects the homogeneous
nucleation of `simple' metastable liquids (for example, LJ and water),
HS colloids and glassy systems. The flow-induced crystallization (FIC)
of polymers is an intriguing nonequilibrium kinetically controlled phase
transition \cite{Graham2009} that is of immense technological importance
\cite{Samon1999, Doufas2000, Zheng2007, Xu2015}. However, the study of
the crystallization of polymers is fraught with considerably greater
complexity, arising from the several levels of structural organization
that can characterize such phase transitions. In Section
\ref{structural-effects-of-shear}, we alluded to how strongly aligned,
elongated nuclei, called "shish-kebabs", form in sheared polymer
melts, which are drastically different from the ellipsoidal nuclei
formed in simple liquids and glassy systems. The difference in morphology is made explicit in Figure \ref{fig:polymerShape}. The crystallization process
of polymer melts differs fundamentally from that in simple liquids due
to the interactions and associations of constituent chains. Figure \ref{fig:polymerShape}(a) pictorially depicts the various addition mechanisms of polymer segments to the nucleus, according to the coarse-grained Graham and Olmsted (GO) \cite{Graham2009, Graham2010a} model.  Entanglement of polymer chains refers to the topological impediment of molecular
motion by other chains \cite{Wool1993, McLeish2002}, which could also be
relevant for nucleation. Here, we briefly outline key conclusions. FIC
is discussed in more detail elsewhere
\cite{Graham2010, Graham2014, Wang2016, Cui2018, Graham2019}.

Model alkane systems are an illustrative example of the intricacies
involved in polymer crystallization. In a study of the sheared
nucleation of short chain alkanes, \(n-\textrm{eicosane}\) (henceforth
referred to as C20) and \(n-\textrm{hexacontane}\) (C60), the
application of a planar shear flow was found to speed up crystallization
and enhance the crystal growth rate \cite{Jabbarzadeh2009}. It was also
concluded that the chain length is directly linked with the magnitude of
the `critical shear rate,' defined as the shear rate above which
crystallization is enhanced by shear. An inference that can be drawn is
that the relaxation time of the molecules is related to the
susceptibility to shear-induced crystallization
\cite{Balzano2008, Jabbarzadeh2009}.

Using NEMD simulations, Anwar et al \cite{Anwar2014}. studied the
quiescent and sheared nucleation of the short chain alkane,
\(n-\textrm{eicosane}\) (C20) and a long chain alkane,
\(n-\textrm{pentacontahectane}\) (C150), at \(20-30\%\) undercooling.
Under quiescent conditions, it was found that nucleation proceeds
according to the same mechanism for both the long and short chains: the
chain segments sequentially first align, straighten, and then coagulate
into clusters. For low shear rates, the authors discovered that both the
long and short chains obey the same nucleation mechanism as that of
quiescent nucleation. Conversely, at high shear rates, a difference in
behaviour was observed for C20 and C150. Chains of C20 simultaneously
align and straighten, which is followed by a local increase in density.
In contrast, C150 chains follow the same sequence of events as those in
quiescent nucleation. These observations are consistent with both
experiments and theory \cite{Coppola2004}.

A recent NEMD study on the crystallization of entangled linear polymer
chains (C1000), consisting of \(1000\) monomers, investigated the
interplay of entanglement, shear and temperature \cite{Anwar2019}. A
more pronounced effect of the imposed shear rates was found on the C1000
chains at higher temperatures. A weak dependence of the induction strain
(defined as the product of the nucleation induction time and shear rate)
on the shear rate was also determined.

Clearly, flow can play a crucial non-trivial role in polymer
crystallization. The generally accepted consensus of the molecular
effects of shear is that an applied flow unravels and stretches polymer
chains, which are initially in a quiescent random-coil configuration,
thereby inhibiting their conformational freedom. This enables the
polymer chains to assemble in the elongated configurations that are
precursors to crystallization. Capturing these events is the primary
challenge of modelling polymer systems in simulations. Brute-force
approaches and rare event biasing techniques, often paired with an appropriate computational model (for
example, the GLaMM model \cite{Graham2003}, and the GO model \cite{Graham2009, Graham2010a}), are generally the
techniques of choice employed. It is not yet clear if a phenomenological
theory (like the CNT-based approaches discussed for simple liquids) can
be suitably modified for polymer crystallization. Although FFS has been
utilized for structural relaxation and translocation of quiescent
polymers
\cite{Huang2008, HernandezOrtiz2009, Ruzicka2012, Cao2015, Zhu2017a, Rezvantalab2018},
determining adequate order parameters for describing the complex process
of sheared polymer crystallization, which typically involve the registry
of polymer chains and elongation of nuclei under flow, could be
especially daunting.

\subsection{\label{experiments}Purview of Experiments}

Water is ubiquitous and widely studied, which we consider first as a
representative member of the class of simple liquids. The homogeneous
quiescent nucleation of ice is often estimated by droplet experiments
\cite{Bauerecker2008, Manka2012, Riechers2013, Sellberg2014, Laksmono2015, Atkinson2016},
wherein the homogeneous nucleation rate is generally equated to the
volume nucleation rate. Nucleation experiments of water are hampered by
the notorious "no-man's land" \cite{Mishima1998}, which sets a
practical limit to the supercooling that can be achieved, due to rapid
ice crystallization that prevents characterization of the phase
transformation. The border no-man's land, situated at around
\(227-228 \ K\), has been penetrated by subversive studies aided by
advanced cutting-edge technologies
\cite{Manka2012, Laksmono2015, Xu2016, Handle2017, Kim2017, Ni2018, Saito2019}.
However, in general, experimentally accessible supercoolings are low or
moderate within the scope of simulations and theory, which perhaps
contributes to the infamous divide between computational results and
experiments \cite{Dixit2001, Sosso2016a, Blow2021}.

In crystallization experiments, shear can be imposed via a Couette cell
\cite{Panine2002}, Taylor-Couette flow system \cite{Liu2013, Liu2014},
Capillary setup \cite{Forsyth2014}, viscometer
\cite{Southern1970a, Southern1970, Crystal1971, Southern1972}, rheometer
\cite{Haas1969}, short-term shearing approaches
\cite{JaneschitzKriegl1990, Kumaraswamy1999, Geng2009, Kornfield2002}
and microfluidics \cite{Stroobants2020}. In general, equipment used in
sheared crystallization experiments can be classified into two
categories: capillary and rotational devices \cite{Stroobants2020}. In a
capillary device, the pressure difference in a capillary is used to
create high shear rates of up to \(3\times10^5 \ s^{-1}\) for short
millisecond length bursts of operation. Unfortunately, a range of shear
rates are imposed instead of a constant shear rate due to the
pressure-driven Poiseuille flow. Rotational devices attempt to generate
a Couette flow but risk creating turbulent flows at higher shear rates
\cite{Bekard2011} and inhomogeneous flows. On the other hand,
microfluidic devices impose flows with low Reynolds numbers (and
consequently impose low shear rates), eliminating the possibility of
turbulent flows \cite{Stroobants2020}. By contrast, Couette-Taylor
devices rely on turbulent flows to enhance mixing \cite{Jung2004}.

The nature of these disparate experimental techniques naturally leads to
the conundrum of how a uniform shear rate can be imposed while
subjecting the system to the supercoolings required for the nucleation
of water, and by extension, other simple liquids. It seems likely that
temperatures within no-man's land will certainly be out of reach of
conventional Couette flow type experiments.

On the other hand, analysing the sheared nucleation of dense
suspensions, colloids, polymers and proteins presents less serious
challenges from the point of view of experiments. Experimental aspects
of FIC of polymers has been extensively reviewed in previous work
\cite{Wang2016, Cui2018}. Recent theoretical results on a colloidal
suspension of PMMA (poly(methyl methacrylate) indicate that colloids,
subjected to a constant linear flow, will exhibit non-monotonicity in
the nucleation rate within shear rates of a few seconds inverse
\cite{Mura2016}. Imposing such linear flows are is well-within the
capabilities of the arsenal of modern experimental techniques, at
ambient temperatures.

\subsection{\label{future-scope-and-challenges}Future Scope and
Challenges}

Brute-force approaches (of which MFPT is primarily used for sheared
homogeneous nucleation) are straightforward in application and can
provide unbiased estimates if sufficiently good statistics are obtained.
Unfortunately, the spatio-temporal resolution of unbiased MD simulations
dictates that obtaining the requisite ensemble of nucleating
trajectories is computationally very expensive. In fact, brute-force
approaches are infeasible for systems which do not nucleate within
achievable simulation times. Furthermore, every imposed shear rate
necessitates generating a separate ensemble of trajectories. For
example, if \(J\) values at \(4\) different shear rates are desired, it
can be surmised that at least \(4\) times the computational cost of
estimating \(J_0\) at the same thermodynamic conditions will be
incurred. Therefore, calculations at moderate and low supercoolings, for
a finite number of shear rates are not yet practically feasible, and
usually only deep supercoolings can be studied using brute-force
approaches.

CNT-based approaches describing sheared homogeneous nucleation are
relatively less expensive, and work reasonably well for several systems
over a wide range of metastabilities. The simple governing expressions
facilitate general analyses and enable overarching inferences to be
drawn for a variety of conditions. However, CNT involves considerable
approximations, even when augmented with seeded simulations and other
methodological improvements. Qualitative agreement notwithstanding,
CNT-based techniques are fundamentally impaired by the very assumptions
that impart this class of methods with their trademark simplicity.

FFS could be an attractive alternative, and is known to provide reliable
insight into microscopic structures, preferred pathways, and molecular
mechanisms. FFS is certainly a promising approach, but though FFS is
`embarrassingly parallel' it remains computationally exorbitant
\cite{Hussain2020}. FFS has the same fatal flaw as brute-force
approaches: a bundle of trajectories must be generated for every shear
rate of possible interest. Perhaps this is why, till date, only the
sheared two-dimensional Ising model has been analysed using FFS, to the
best of our knowledge \cite{Allen2008, Allen2008a}. A viable strategy
might be to pair FFS with other less costly computational approaches to
identify shear regimes of interest.

Another point illuminated by recent work on water
\cite{Luo2020, Goswami2020a, Goswami2021a} is that the optimal shear
rates, at moderate and deep supercoolings, are several orders of
magnitude greater than shear rates that can be imposed by experiments.
Water is an exemplary example: generating a constant linear velocity
profile within no-man's land may not be feasible in experiments. Adding
to this difficulty is the increased uncertainty of simulation results
for lower shear rates which are experimentally achievable
\cite{Luo2020}. This may be one of the contributing factors for why the
non-monotonic behaviour of \(J\) with \(\dot{\gamma}\), predicted by
simulations and theory, has not yet been observed in experiments of
simple Newtonian liquids.

Shear can cause a plethora of interesting effects in polymers melts and
glasses, including jamming transitions, annealing, shear-induced
ordering etc. Brute-force approaches and coarse-grained models are
typically used for analysing the effects of flow on polymer phase
transitions. It remains to be seen whether CNT pathways and the
theoretical models developed for simple liquids can be applied to study
the complex rheological behaviour of polymer melts.

Although we have not discussed sheared heterogeneous nucleation in this
review, a recent study on mW water showed that the sheared heterogeneous
nucleation rates are also non-monotonic with shear \cite{Luo2019}.
Unbiased MD simulations were used by the authors, but brute-force
approaches are not always feasible, even for heterogeneous nucleation
\cite{Sosso2016a}. On the other hand, it could be possible to extend
CNT-based approaches to heterogeneous nucleation, which would presumably
be applicable to a wider variety of systems and surfaces. Heterogeneous
CNT methods, relying on random structure searches \cite{Pedevilla2018}
and machine learning \cite{Fitzner2020}, have been proposed and employed
for quiescent nucleating systems. Though the prospect of a CNT-based
approach for sheared heterogeneous nucleation is appealing, seemingly
insurmountable challenges are also involved. A rigorous theoretical
treatment of the exotic flow field around a nucleus sitting on a surface
may not be tractable. In addition, while the internal structures and
polymorphic diversity of nuclei are usually ignored for homogeneous
nucleation, the microstructure may be significant for heterogeneous
nucleation.

It is abundantly clear that the sheared homogeneous nucleation of
Newtonian liquids, dense suspensions, glassy systems, and polymer melts
exhibits richly diverse and complex phase behaviour. Theory and
simulations can be particularly relevant for studying the elusive
sheared homogeneous nucleation of simple liquids and glassy systems,
especially for high shear rates which are beyond the current
capabilities of Couette flow inducing experiments. We hope that
computational techniques and theoretical tools will be developed
further, striving towards greater fidelity with available experimental
data and contributing to our understanding of the fascinating phenomenon
of nucleation under flow.


\hypertarget{acknowledgements}{%
\section{Acknowledgements}\label{acknowledgements}}

This work was supported by the Science and Engineering Research Board
(sanction number STR/2019/000090 and CRG/2019/001325). 

\printbibliography


\end{document}